\documentclass[%
 aip,
 amsmath,amssymb,
 reprint,%
]{revtex4-1}

\usepackage{graphicx}
\usepackage{dcolumn}
\usepackage{bm}
\usepackage{lineno}

\usepackage[utf8]{inputenc}
\usepackage[T1]{fontenc}
\usepackage{mathptmx}
\usepackage{xcolor}
\usepackage{adjustbox}
\usepackage{multirow} 
\usepackage{ulem}

\usepackage{setspace}

\begin{document}

\preprint{Preprint}

\title[Asymmetrical voltage response in neurons]{Asymmetrical voltage response in resonant neurons shaped by nonlinearities}

\author{R. F. O. Pena}
\email{pena@njit.edu}
\affiliation{Federated Department of Biological Sciences, New Jersey Institute of Technology and Rutgers University, Newark, New Jersey, NJ 07102, USA
}%
\affiliation{Institute for Brain and Neuroscience Research, New Jersey Institute of Technology, Newark, NJ 07102, USA}
\affiliation{
Department of Physics, School of Philosophy, Sciences and Letters of Ribeir\~ao Preto, University of S\~ao Paulo, CEP 14040-901, Ribeir\~ao Preto, Brazil}
\affiliation{These authors have contributed equally to this work.}

\author{V. Lima}%
\affiliation{
Department of Physics, School of Philosophy, Sciences and Letters of Ribeir\~ao Preto, University of S\~ao Paulo, CEP 14040-901, Ribeir\~ao Preto, Brazil}
\affiliation{These authors have contributed equally to this work.}
 
\author{R. O. Shimoura}
\affiliation{
Department of Physics, School of Philosophy, Sciences and Letters of Ribeir\~ao Preto, University of S\~ao Paulo, CEP 14040-901, Ribeir\~ao Preto, Brazil}

\author{C. C. Ceballos}
\affiliation{
Department of Physics, School of Philosophy, Sciences and Letters of Ribeir\~ao Preto, University of S\~ao Paulo, CEP 14040-901, Ribeir\~ao Preto, Brazil}
\affiliation{Department of Physiology, School of Medicine of Ribeir\~ao Preto, University of S\~ao Paulo, CEP 14049-900, Ribeir\~ao Preto, Brazil}

\author{H. G. Rotstein\footnote{Graduate Faculty, Behavioral Neurosciences (BNS) Program, Rutgers University, USA}\footnote{Corresponding Investigator, CONICET, Argentina}}
\email{horacio@njit.edu}
\affiliation{Federated Department of Biological Sciences, New Jersey Institute of Technology and Rutgers University, Newark, New Jersey, NJ 07102, USA
}%
\affiliation{Institute for Brain and Neuroscience Research, New Jersey Institute of Technology, Newark, NJ 07102, USA}

\author{A. C. Roque}
\email{antonior@usp.br}
\affiliation{
Department of Physics, School of Philosophy, Sciences and Letters of Ribeir\~ao Preto, University of S\~ao Paulo, CEP 14040-901, Ribeir\~ao Preto, Brazil}

\date{\today}

\begin{abstract}
\textbf{Abstract}
The conventional impedance profile of a neuron can identify the presence  of resonance and other properties of the neuronal response to oscillatory inputs, such as nonlinear response amplifications, but it cannot distinguish other nonlinear properties such as asymmetries in the shape of the voltage response envelope.
Experimental observations have shown that the response of neurons to oscillatory inputs preferentially enhances either the upper or lower part of the voltage envelope in different frequency bands. These asymmetric voltage responses arise in a neuron model when it is submitted 
to high enough amplitude oscillatory currents of variable frequencies. We show how the nonlinearities associated to different ionic currents or present in the model as captured by its voltage equation lead to asymmetrical response and how high amplitude oscillatory currents emphasize this response. We propose a geometrical explanation for the phenomenon where asymmetries result not only from nonlinearities in their activation curves but also from nonlinearites captured by the nullclines in the phase-plane diagram and from the system's time-scale separation. In addition, we identify an unexpected frequency-dependent pattern which develops in the gating variables of these currents and is a product of strong nonlinearities in the system as we show by controlling such behavior by manipulating the activation curve parameters. The results reported in this paper shed light on the ionic mechanisms by which  brain embedded neurons process oscillatory information.
\end{abstract}

\maketitle


\begin{quotation}

The relationship between a stimulus and the neuron's response is important for information transmission in neuronal systems. In response to an oscillatory stimulus, some neurons exhibit subthreshold (membrane potential) resonance where the steady state voltage amplitude response is maximal at a preferred (resonant) frequency. This is typically measured by the impedance vs. frequency curve (impedance profile). The use of the impedance for this purpose implicitly assumes that the upper and lower voltage envelope (curves of the maximal and minimal steady state voltages as a function of the input frequency) are symmetrical (or quasi-symmetrical) with respect to the baseline (holding) potential. However, because of the presence of nonlinearities, primarily in the voltage dependencies, this assumption is valid only for low enough input amplitudes. In fact, asymmetric voltage responses have been experimentally observed in various neuronal systems. How do the systems' nonlinearities shape the neuronal voltage response to oscillatory inputs remains an open question. In this paper we address this issue by using modeling, dynamical systems tools and numerical simulations. We analyze the role played by the nonlinear intrinsic properties of neurons and time scales, primarily imposed by the ionic currents, in shaping the upper and lower voltage response envelopes by discussing a number of representative cases. Our results contribute to the understanding of the resonant properties of neurons and their role for information processing.

\end{quotation}

\section{\label{sec:level1}Introduction}

Under the right combination of ionic currents and input frequency, neurons can exhibit subthreshold resonance, i.e. the membrane voltage response  peaks at a non-zero frequency of the oscillatory input current. Experimental and theoretical studies have related this phenomenon to  interacting processes at the ionic level, e.g. time constants and conductances \cite{hutcheon1996, hutcheon2000,rotstein2014,pena2018}. However, there are still open questions such as whether and, if yes, how resonance has any functionality for signal processing \cite{chacrom2001,remme2014}, and how alterations in the biophysical properties of neurons modify the resonant properties of the voltage response \cite{rotstein2017n}.

\begin{figure*}
    \centering
    \includegraphics[scale=0.34]{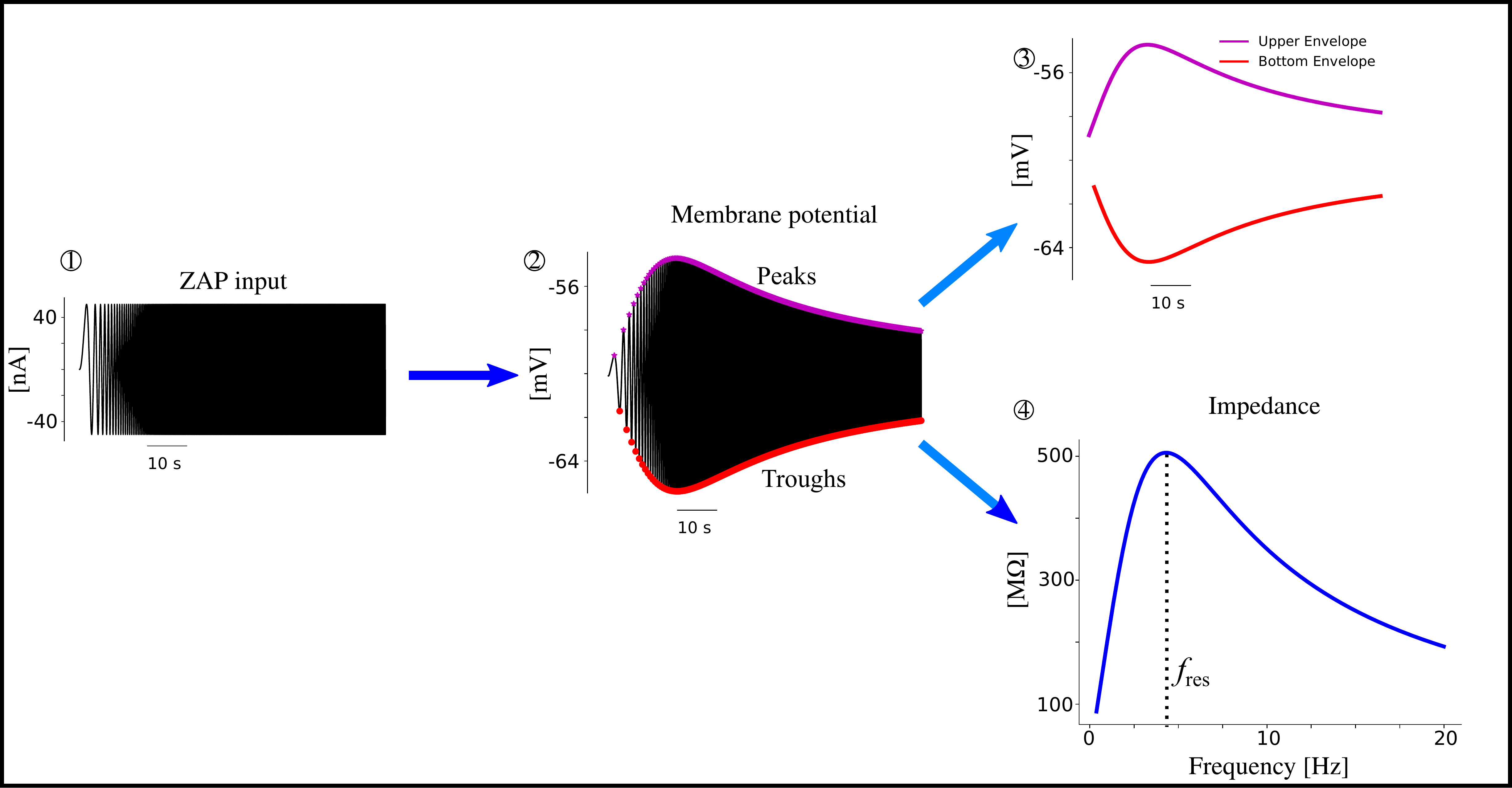}
    \caption{Typical behavior of a neuron upon stimulation with a ZAP input. 1: Constant amplitude input with a linearly swept frequency (ZAP input). 2: Voltage response of a neuron upon ZAP stimulation where a proper algorithm detects peaks and troughs from the voltage series. 3: Peaks and troughs connected are used to capture the upper and bottom envelopes. 4: Impedance profile is built with the same information of peaks and troughs. From the impedance profile, it is possible to obtain the resonance frequency $f_{\rm res}$.}
    \label{fig:illustration}
\end{figure*}

Resonant currents such as the hyperpolarization-activated current ($I_{\rm h}$) contribute to the emergence of resonance. Recent studies have established the role that $I_{\rm h}$ plays in determining the properties of neuronal resonance, based on (i) dynamical systems analysis \cite{rotstein2017, zhao2018}; (ii) characterization of response curves, mostly measured in terms of the impedance ($Z$) magnitude and phase \cite{hutcheon2000,sanhueza2014,pena2018}; and (iii) autocorrelograms \cite{sanhueza2005}. Another resonant current is the M-current ($I_{\rm M}$) which produces comparable
effects to $I_{\rm h}$. In addition to the resonant currents, amplifying currents such as the potassium inward rectifier ($I_{\rm Kir}$) and persistent sodium ($I_{\rm NaP})$ do not create resonance but generate positive feedback effects, which in turn can augment the resonant effects when present~\cite{hutcheon2000,rotstein2015}.

Typically, by applying an oscillatory current with constant amplitude and linearly increasing frequency, such as the ZAP (Impedance Amplitude Profile) function\cite{cali2008}, and measuring the voltage response, one can construct the input-output relationship over a range of different input frequencies. The upper and lower voltage envelope-curves can be used to demonstrate this relationship by simply connecting voltage peaks and voltage troughs as frequency increases. In the frequency domain, the impedance amplitude profile of the neuron identifies this property and is given by the ratio of the Fourier transforms of the output and the input \cite{hutcheon2000,rotstein2014}.  The resonance frequency $f_{\rm res}$ is the impedance profile peak frequency (see illustration in Fig.~\ref{fig:illustration}). Nevertheless, the impedance profile, traditionally used to measure the response of neurons to oscillatory inputs, does not necessarily capture the properties of the neuronal response, which may require looking back into the envelope responses.
When the ZAP input is applied with low amplitudes (tens of pA or less) and because the dynamics is approximately linear, in general, the voltage response oscillations are symmetric about a reference voltage line, i.e. the peak frequency of the upper voltage envelope coincides with the trough frequency of the lower voltage envelope, justifying the use of the impedance profile.

The steady state response of linear systems to oscillatory inputs is characterized by three properties: (i) the number of input and output cycles coincide and the output amplitude is uniform across cycles for a given input frequency, (ii) the output amplitude is proportional to the input amplitude rendering the impedance independent of the input amplitude, and (iii) the output is symmetric with respect to the stable equilibrium of the unforced system around which the forced system oscillates. In this work we are implicitly assuming (i), while (ii) and (iii) are violated. In other words, in this work we are considering nonlinear models for which (i) is satisfied, while (ii) and (iii) are not. 

However, for voltage-dependent ionic currents (e.g., $I_{\rm h}$, $I_{\rm M}$) it is expected that large voltage changes bias the impedance response differently between depolarized and hyperpolarized currents. 
An amplitude high enough to activate these currents accentuates  nonlinearities leading to a neuron's response with asymmetrical upper and lower voltage envelopes. In fact, recent work reported evidence of asymmetric voltage responses to ZAP functions, where the peak frequency of the upper voltage envelope differs from the trough frequency of the lower voltage envelope \cite{tohidi2009,schreiber2009,rotstein2014b,rotstein2015,fischer2018}.

The question arises of how do the intrinsic nonlinear neuronal properties shape  the voltage response (upper and lower envelopes) and the impedance profile. To address these issues, in this work we investigate the asymmetric properties of the voltage response of neurons to oscillatory inputs and how these properties are shaped by the participating ionic currents. However, because the impedance profile does not necessarily capture asymmetries in the upper and lower envelopes, we propose a different version of the conventional impedance by looking at the upper and lower voltage envelope profiles normalized by the input amplitude, which we refer to as the upper ($Z^+$) and lower impedances ($Z^-$). For linear systems, the impedance is the difference between these two normalized quantities. We show how the upper and lower impedances may be used to characterize asymmetrical responses in the biophysical model. Our simulations are constrained within the biological range  \cite{pospischil2008}. We explain how asymmetrical responses are related to the activation curves of the modeled ionic currents, in particular we show that the gating variables of these currents may lead to an unexpected frequency-dependent pattern. We explain our results by using dynamical systems tools (phase-plane diagrams). Finally, we use a simplified piecewise linear system subject to an oscillatory input to explain our findings in a more tractable way.

\section{Methods}

\subsection{Model}

We use a conductance based neuron model with voltage-dependent currents and a leak current. The membrane potential ($V$) of the model obeys 

\begin{equation}
C_\text{m} \frac{dV(t)}{dt} = -\sum_iI_{i}(V,t) - I_\text{leak}(V) + I_\text{ZAP}(t) + I_{\rm DC}, 
\label{eq:HH}
\end{equation}

\noindent where $C_{\rm m}=1$ $\mu$F/cm$^2$ is the membrane capacitance and $I_{\rm leak}$ the leak current, given by $I_{\rm leak}=g_{\rm leak}(V-E_{\rm leak})$ where $g_{\rm leak}$ is the leak conductance and $E_{\rm leak}$ the reversal potential of the leak current. We chose the geometry of a cylinder with 70 $\mu$m of diameter and 70 $\mu$m of length which condenses soma and dendrite in a single compartment preserving the average capacitance of a pyramidal cell ($C \approx 150$~ pF) \cite{pena2018,tamagnini2015}. In the case of other ionic currents, $I_i$ may represent the hyperpolarization-activated current ($I_{\rm h}$), the M-current ($I_{\rm M}$), the potassium inward rectifier ($I_{\rm Kir}$), or the persistent sodium ($I_{\rm NaP})$. They follow the Hodgkin-Huxley formalism \cite{hodgkin1952} with dynamics given by 

\begin{equation}
I_{i} = \bar{g}_{i}A_{i}(V,t)(V-E_{i}), 
\end{equation}

\noindent where $\bar{g}_{i}$ is the maximal conductance and $E_i$ the reversal potential. $A_{i}(V,t)$ is the activation variable defined as

\begin{equation}
\frac{dA_{i}(V,t)}{dt}= \frac{A_{i}^{\infty}(V)-A_{i}(V,t)}{\tau_{i}},
\label{eq:Ah}
\end{equation} 

\noindent where $\tau_{i}$ is the time constant of the activation variable, and $A_{i}^{\infty}$ is the asymptotic value of the variable, which follows the Boltzmann formalism  

\begin{equation}
A_{i}^{\infty}(V) = \frac{1}{1+\exp\left( \frac{s(V-V_{1/2})}{k} \right)}, 
\label{eq::a_inf}
\end{equation}

\noindent where the values $V_{1/2}$ and $k$ control the sigmoid function: the first is the voltage value for which the ionic current is half activated (i.e., $A_i(V_{1/2})=0.5$) and the second determines the slope of the sigmoid function. The sign of $s$ controls whether it activates with hyperpolarization or depolarization.  $I_{\rm ZAP}$ represents the ZAP current, as considered in Ref.~\onlinecite{cali2008}, given by 

\begin{align}
\begin{cases}
I_\text{ZAP} = A_{\rm in}\sin[\pi (f(t)-F_\text{start})(t-t_\text{start})], \\
f(t) = F_\text{start} + (F_\text{stop}-F_\text{start})(t-t_\text{start})/(t_\text{stop}-t_\text{start}),
\end{cases}
\label{eq:ZAP}
\end{align}

\noindent where $F_{\rm start}$ ($F_{\rm stop}$) is the initial (final) frequency of the $I_{\rm ZAP}$, and $t_{\rm start}$ ($t_{\rm stop}$) is the initial (final) time limit. The amplitude of the ZAP current ($I_{\rm ZAP}$) is given by $A_{\rm in}$, which will be shown in every figure.

A constant current was applied to keep the neuron at different $V_{\rm hold}$ values ($I_{\rm DC}$). Values of $V_{\rm hold}$ will be indicated in the results section. All parameters used are displayed in Table \ref{Tab::parameters}.

\begin{table*}[!htb]
	\centering
    \begin{adjustbox}{width=0.8\textwidth}
    \begin{tabular}{|c|c|c|c|c|c|c|}
    \hline 
    \multicolumn{7}{|c|}{\textbf{Neuron model parameters}}\tabularnewline
    \hline 
    \textbf{Current type} & \textbf{$E$ {[}mV{]}} & \textbf{$\bar{g}$ {[}S/${\rm cm^{2}}${]}} & \textbf{$\tau$ {[}ms{]}} & \textbf{$s$} & \textbf{$V_{1/2}$ {[}mV{]}} & \textbf{$k$ {[}mV{]}}\tabularnewline
    \hline 
    \textbf{Leak} & -90 & $6.56\times10^{-5}$ & --- & --- & --- & ---\tabularnewline
    \hline 
    \textbf{$I_{{\rm h}}$} & -30 & $6.56\times10^{-5}$ & $\in[10;1000]$ & 1 & -$82$ & $9$\tabularnewline
    \hline 
    \textbf{$I_{{\rm M}}$} & -30 & $6.56\times10^{-5}$ & 100 & -1 & -82 & 9\tabularnewline
    \hline 
    \textbf{$I_{{\rm NaP}}$} & 50 & $2.0\times10^{-5}$ & Eq.~(\ref{eq:tau_Inap}) & -1 & -48 & 10\tabularnewline
    \hline 
    \textbf{$I_{{\rm Kir}}$} & -100 & $5.76\times10^{-5}$ & Eq.~(\ref{eq:tau_Ikir}) & 1 & -98.92 & 10.89\tabularnewline
    \hline 
    \multicolumn{7}{|c|}{\textbf{Stimulation parameters}}\tabularnewline
    \hline 
    \multirow{2}{*}{ZAP} & \textbf{$F_{{\rm start}}$ {[}Hz{]}} & \textbf{$F_{{\rm stop}}$ {[}Hz{]}} & \textbf{$t_{{\rm start}}$ {[}s{]}} & \multicolumn{3}{c|}{\textbf{$t_{{\rm stop}}$ {[}s{]}}}\tabularnewline
    \cline{2-7} \cline{3-7} \cline{4-7} \cline{5-7} \cline{6-7} \cline{7-7} 
     & 0.001 & 20 & 2 & \multicolumn{3}{c|}{620}\tabularnewline
    \hline 
    \textbf{$I_{{\rm DC}}$} & \multicolumn{6}{c|}{keeps  $V=V_{{\rm hold}}$ {[}mV{]}}\tabularnewline
    \hline 
    \end{tabular}
    \end{adjustbox}
	\caption{\label{Tab::parameters} Model parameters used in this work.}
\end{table*}

The $I_{\rm NaP}$ time constant ($\tau_{\rm NaP}$ in [ms]) is different for values of $V$ higher and below $-40$~mV, as considered in Ref.~\onlinecite{traub2003}, which creates a fast/slow activation depending on the voltage, and is described as

\begin{align}
\tau_{\rm NaP}=
\begin{cases}
0.025+0.14\exp((V+40)/10) & \text{if } V\leq-40 \text{ mV} \\
0.02+0.145\exp((-V-40)/10) & \text{otherwise. } 
\end{cases}
\label{eq:tau_Inap}
\end{align}

The $I_{\rm Kir}$ time constant ($\tau_{\rm Kir}$ in [ms]) is described as

\begin{align}
\tau_{\rm Kir} = \frac{1}{a\exp(-V/V_{1/2})+b\exp(V/V_{1/2})},
\label{eq:tau_Ikir}
\end{align}

\noindent where $a=6.1$/s and $b=81.8$/s \cite{stegen2011,yim2015}. The $I_{\rm h}$ and $I_{\rm M}$ time constants are displayed in Table~\ref{Tab::parameters}. Note that the time constants for $I_{\rm h}$ and $I_{\rm M}$ are usually larger (slow dynamics) than the values assumed for $I_{\rm NaP}$ and $I_{\rm Kir}$ which are small (fast dynamics). In addition, note that in order to simplify comparison between $I_{\rm h}$ and $I_{\rm M}$ we chose to model the $I_{\rm M}$ current with equal parameters as $I_{\rm h}$ but instead of an observable activation it deactivates with hyperpolarization, i.e. $s=-1$ instead of $s=1$ (see Fig.~\ref{Fig::figCurrents}(a) for examples of activation and deactivation).

\subsection{Measures to identify asymmetries}

Subthreshold resonance in neurons is usually approached by identifying a peak in the impedance magnitude, which is calculated as the ratio of the Fourier transforms of the output voltage and the input current $Z(f)=\text{FFT}_\text{out}/\text{FFT}_\text{in}$. For nonlinear systems, where the output is periodic and has frequency following the input, the impedance magnitude expression can be written as

\begin{align}
    Z(f)=\frac{V_{\max }(f)-V_{\min }(f)}{2 A_{\mathrm{in}}},
    \label{Eq:impedance}
\end{align}

where $V_{\max }(f)$ and $V_{\min }(f)$ are the maximum and minimum voltages obtained from the output voltage for a given frequency $f$ \cite{rotstein2014,rotstein2017n}.

\begin{figure}[!htp]
	\centering
	\includegraphics[scale=0.33]{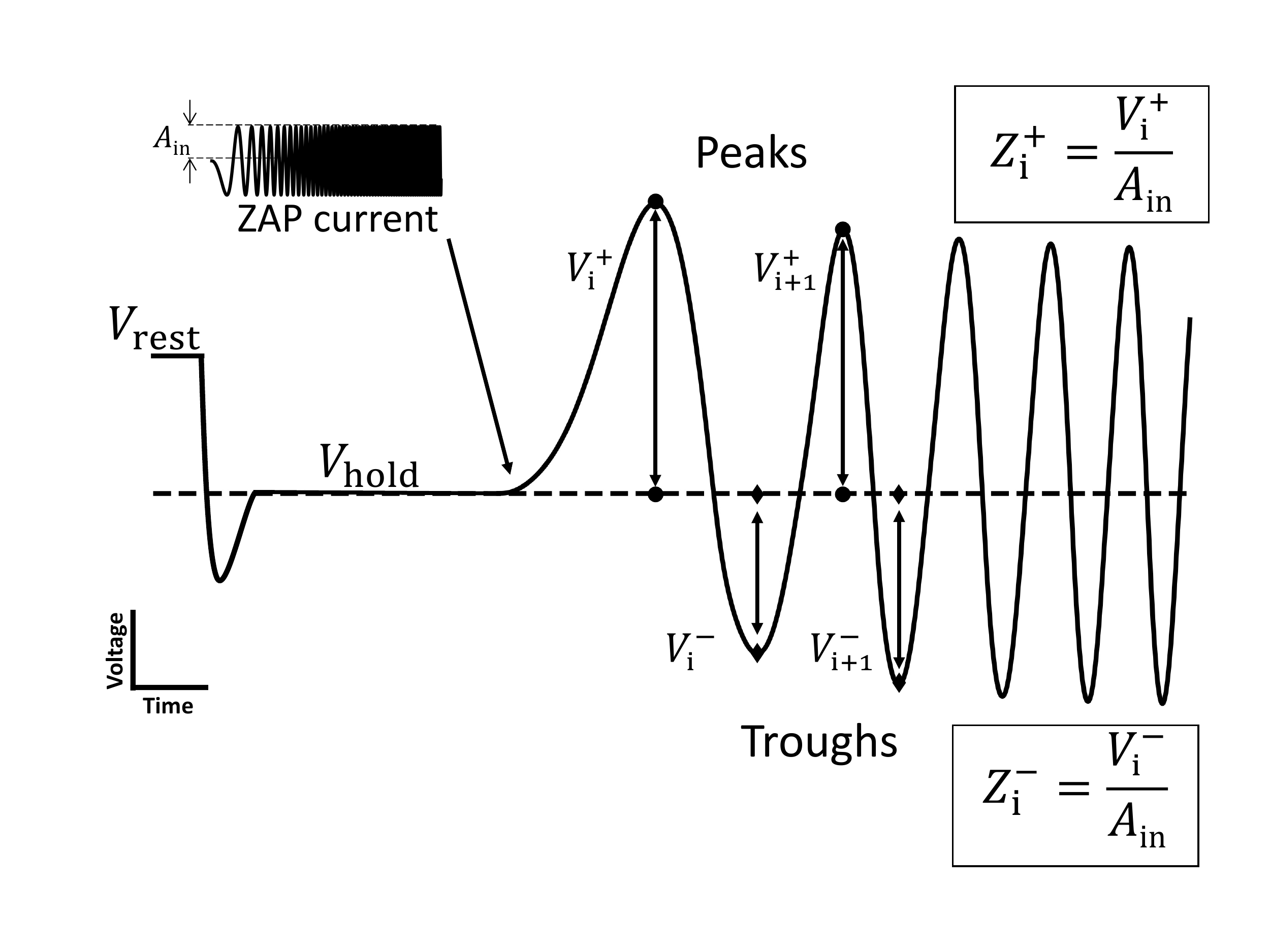}
	\caption{Scheme showing the quantification of $Z^{+}(f)$ and  $Z^{-}(f)$. We show the method for calculation of $Z^{+}(f)$ and $Z^{-}(f)$. At the beginning of the simulation the neuron is at the resting potential ($V_{\rm rest}$). Shortly after that, the potential changes to the holding potential ($V_{\rm hold}$), which is taken as reference (voltages above it are positive and voltages below it are negative). After the ZAP current is applied, the peaks $V^{+}(f)$ and troughs $V^{-}(f)$ of the voltage response are taken, and $Z^{+}(f)$ and $Z^{-}(f)$ are calculated as these respective quantities normalized by the ZAP current amplitude $A_{\mathrm{in}}$ (see text).
	}
	\label{Fig::fig1}
\end{figure}

Here, in order to characterize the asymmetric membrane potential responses we propose two alternative measures to the usual ratio of Fourier transforms. We take the holding potential ($V_{\rm hold}$) as a reference voltage line, so that voltages above it will be positive and voltages below it will be negative, and, for each frequency $f$, measure the magnitudes of the maximum (peak $V^+$) and minimum (trough $V^-$) voltage traces normalized by the ZAP current amplitude ($A_{\rm in}$) (Fig.~\ref{Fig::fig1}). 
We refer to these two quantities, which depend on the frequency $f$ and have dimensions of impedance as upper impedance ($Z^{+}(f)$) and lower impedance ($Z^{-}(f)$) (Eq.~(\ref{Eq:Z_more_less})). These measures stress the presence of asymmetries better than impedance profiles because they do not average upper and lower responses. They are also better than observations of the upper and lower envelopes because they allow a better comparison by highlighting differences due to the normalization.

\begin{align}
Z^{+/-}_i = \frac{V^{+/-}_i}{A_{\mathrm{in}}} \text{, } i=1,2,\dots,N^{+/-},
\label{Eq:Z_more_less}
\end{align}

\noindent where $V^{+/-}$ is the absolute peak/trough distance from $V_{\rm hold}$, and $N^{+/-}$ are the number of peaks/troughs in the time series, respectively. For each peak/trough $i$ we check the corresponding frequency on the ZAP current (from Eq.~(\ref{eq:ZAP})) and use it to draw $Z^{+/-}(f)$. In practice, to obtain $Z^{+/-}(f)$ we simply interpolate the set $\{ Z^{+/-} (f_i)\}$. Notice that the impedance $Z(f)$ is an average over $Z^+(f)$ and $Z^-(f)$.

All simulations were run in the NEURON simulator using the Python interface \cite{hines1997}. Phase-plane analysis was done using MATLAB (The Mathworks, Natick, MA).

\section{Results}

\subsection{Dependencies of asymmetrical subthreshold resonance}\label{sect:dependencies}

Here we show how asymmetrical subthreshold resonance emerges in the  model when a single voltage-dependent (resonant) current is present. We  start with the $I_{\rm h}$ current and explore how the interplay of its time constant and the membrane potential shapes asymmetries on the impedance profile.

\begin{figure}[!htb]
	\centering
	\includegraphics[scale=0.35]{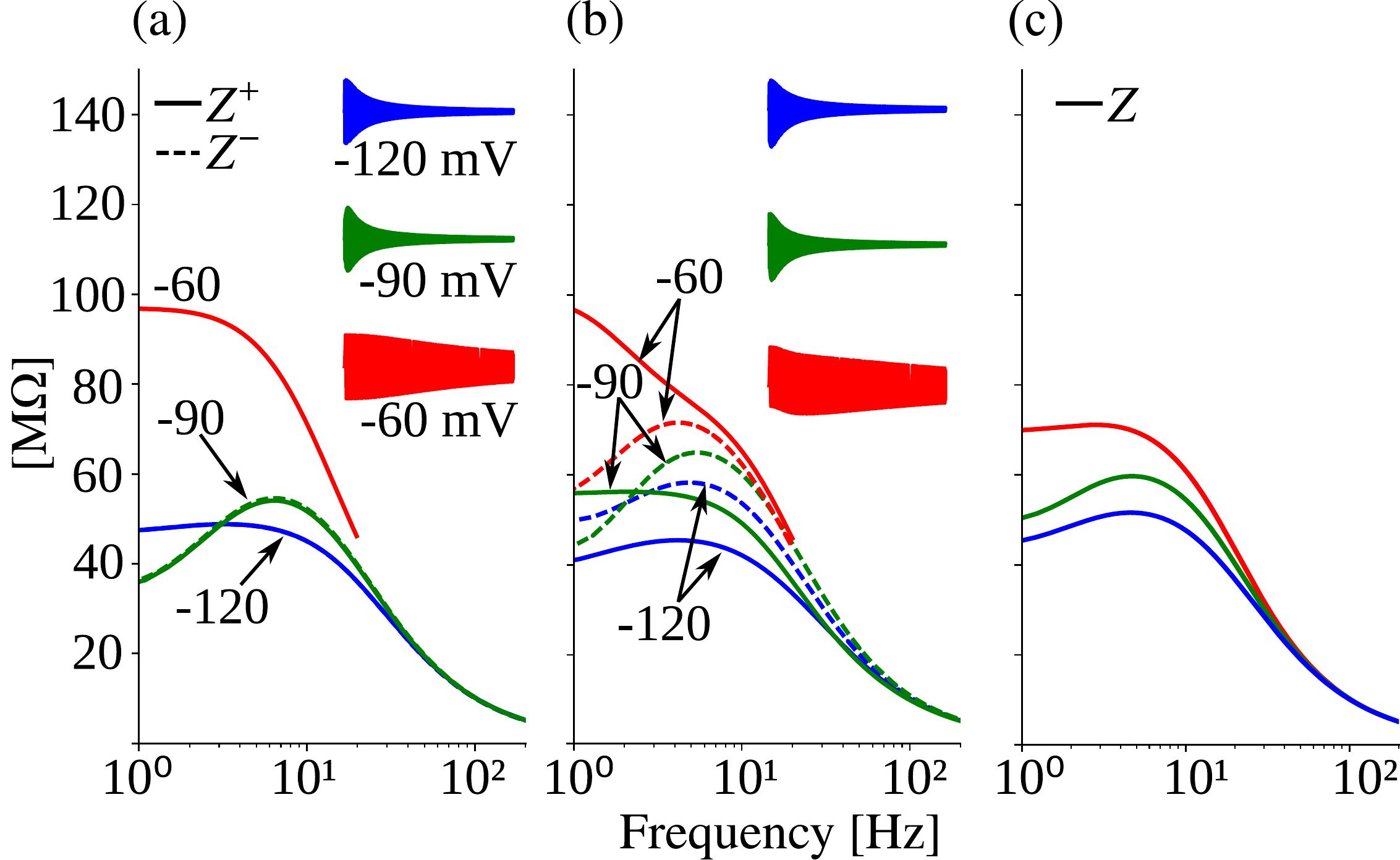}
	\caption{$Z^{+}(f)$ and $Z^{-}(f)$ curves for different ZAP current amplitudes and holding potential values. In these simulations $\tau_{\rm h} = 100$~ms. Solid lines show $Z^{+}(f)$ and dashed lines show $Z^{-}(f)$. Arrows indicate the different values of holding potential: $-120$ mV (blue), $-90$ mV (green) and $-60$ mV (red). Two ZAP current amplitudes were tested: (a) $10$ pA or (b) $1$ nA. Insets show schematic voltage responses to ZAP currents. (c) Impedance profile calculated as in Eq.~(\ref{Eq:impedance}) for the same simulations as in (b).}
	\label{Fig::fig2}
\end{figure}

In Fig.~\ref{Fig::fig2} we show examples of $Z^{+}(f)$ (solid lines) and $Z^{-}(f)$ (dashed lines) computed from our simulations. For the low ZAP current amplitude ($A_{\rm in}= 10$ pA; Fig.~\ref{Fig::fig2}(a)), $Z^{+}(f)$ and $Z^{-}(f)$ are identical. On the other hand, for the high ZAP current amplitude ($A_{\mathrm{in}} = 1$ nA; Fig.~\ref{Fig::fig2}(b)), $Z^{+}(f)$ and $Z^{-}(f)$ display different resonance peaks. Moreover, depending on the holding potential, a resonance peak may exist in $Z^{-}(f)$, but not in $Z^{+}(f)$. 

In Fig.~\ref{Fig::fig2}(c) we present the impedance profiles ($Z(f)$) computed for the same simulations in  Fig.~\ref{Fig::fig2}(b) (colors follow the same scheme). For $V_{\rm hold}=-60$~mV, $Z(f)$ does not display resonance even though in Fig.~\ref{Fig::fig2}(b) $Z^-$ has a resonance peak. On the other hand, for $V_{\rm hold}=-90$~mV both $Z(f)$ and $Z^-$ have resonance peaks. Since $Z(f)$ is a combination of $Z^+$ and $Z^-$, it may fail to capture information about hyperpolarized and depolarized voltages.

The previous examples (Fig.~\ref{Fig::fig2}) suggest that $Z^{+}(f)$ and $Z^{-}(f)$ impedance curves depend on the biophysical properties of $I_{\rm h}$, more specifically $\tau_{\rm h}$ and $V_{\rm hold}$. To study the combined effect of these parameters on the impedance profile (low or band-pass filter), we characterized all four possible scenarios: (i) both $Z^{+}$ and $Z^{-}$ do not exhibit resonance, i.e, are low-pass filters; (ii) $Z^{+}$ is a low-pass filter and $Z^{-}$ is a band-pass filter; (iii) both $Z^{+}$ and $Z^{-}$ are band-pass filters; (iv) $Z^{+}$ is a band-pass filter and $Z^{-}$ is a low-pass filter, which was not detected in our simulations. 

\begin{figure}[!ht]
	\centering
	\includegraphics[width=11cm,height=11cm,keepaspectratio]{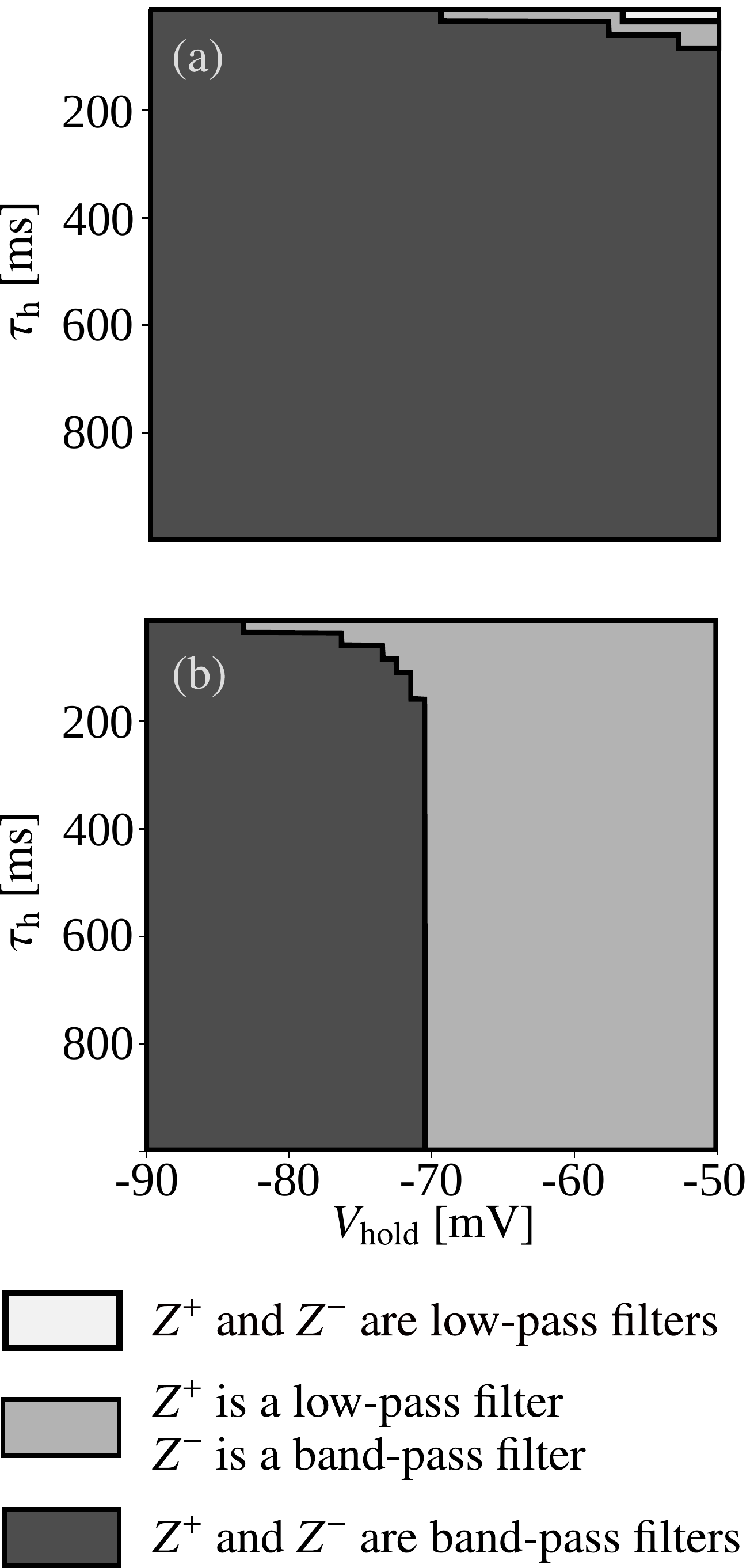}
	\caption{Low-pass and band-pass regions indicated in the two--dimensional diagram spanned by $V_{\rm hold}$ (abscissas axis) and $\tau_{h}$ (ordinates axis), $V_{\rm hold}$--$\tau_{\rm h}$ diagram, for each parameter combination. (a) $A_{\mathrm{in}} = 0.1$~nA, and (b) $A_{\mathrm{in}} = 0.5$~nA.}
	\label{Fig::fig4}
\end{figure}

\par A high ZAP current amplitude abolishes resonance in $Z^{+}$ for depolarized $V_{\rm hold}$ (Fig.~\ref{Fig::fig4}(b)). Thus low-pass or band-pass filtering behavior for depolarizing current is dependent on the current amplitude. This suggests that in a physiological context, the arrival of  oscillatory synaptic inputs might be able to modulate the post-synaptic neuronal response such that high amplitude inputs are preferentially transmitted at low frequencies (a selected frequency band)  if the post-synaptic neuron is depolarized (hyperpolarized). 

\begin{figure}[!htb]
	\centering
	\includegraphics[width=9cm,height=7cm,keepaspectratio]{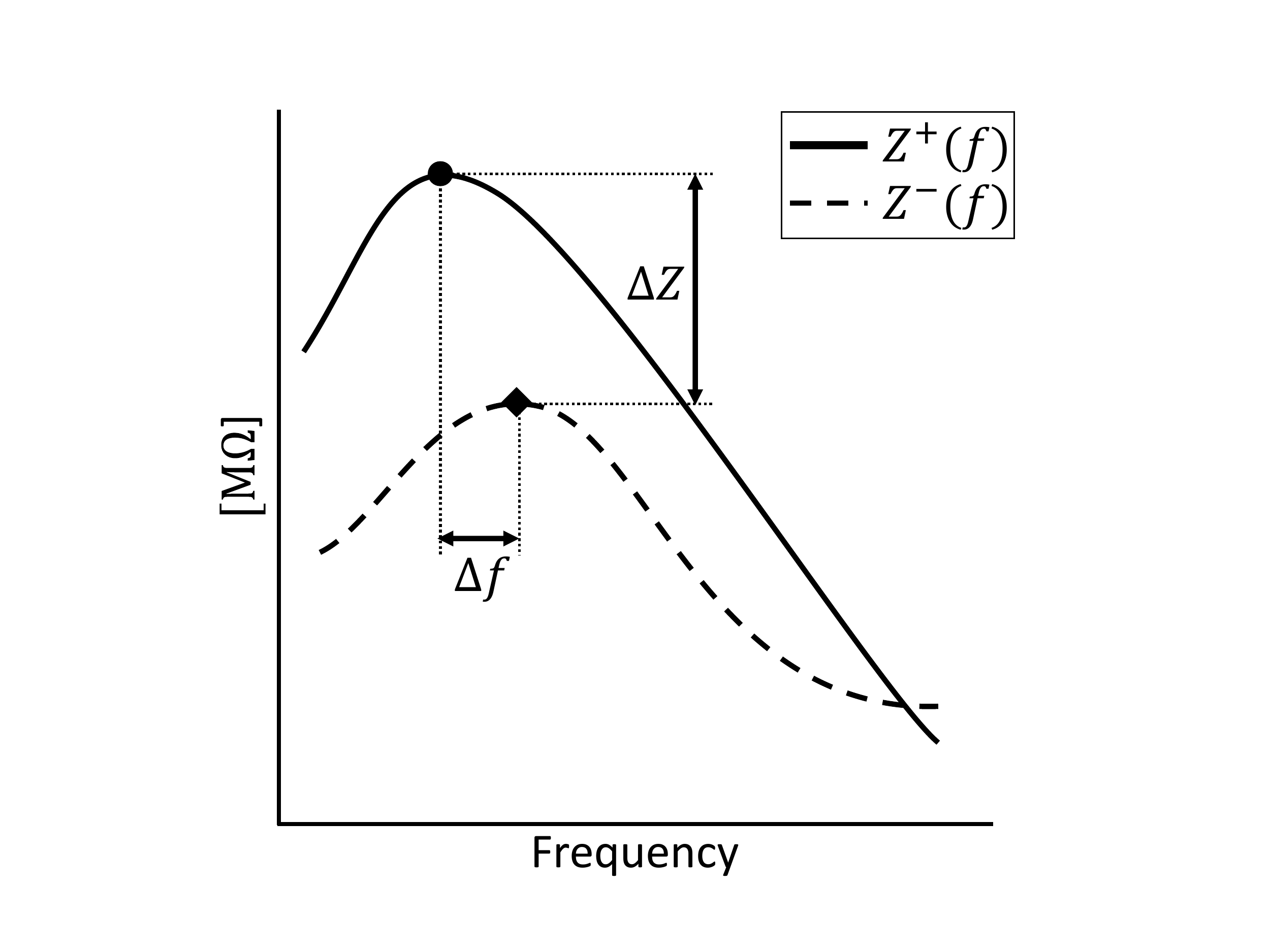}
	\caption{Determination of the shifts $\Delta Z$ and $\Delta f$. The resonance peak shift ($\Delta Z$) and resonance frequency shift ($\Delta f$) definitions. The figure corresponds to ZAP current amplitude $A_{\rm in} = 1$~nA and $V_\text{hold} = -94$~mV. The resonance peak of $Z^+(f)$ is indicated by a filled circle and the resonance peak of $Z^-(f)$ is indicated by a filled diamond. 
	}
	\label{Fig::fig1A}
\end{figure}

\begin{figure*}[!htb]
	\centering
	\includegraphics[width=16cm,height=8cm,keepaspectratio]{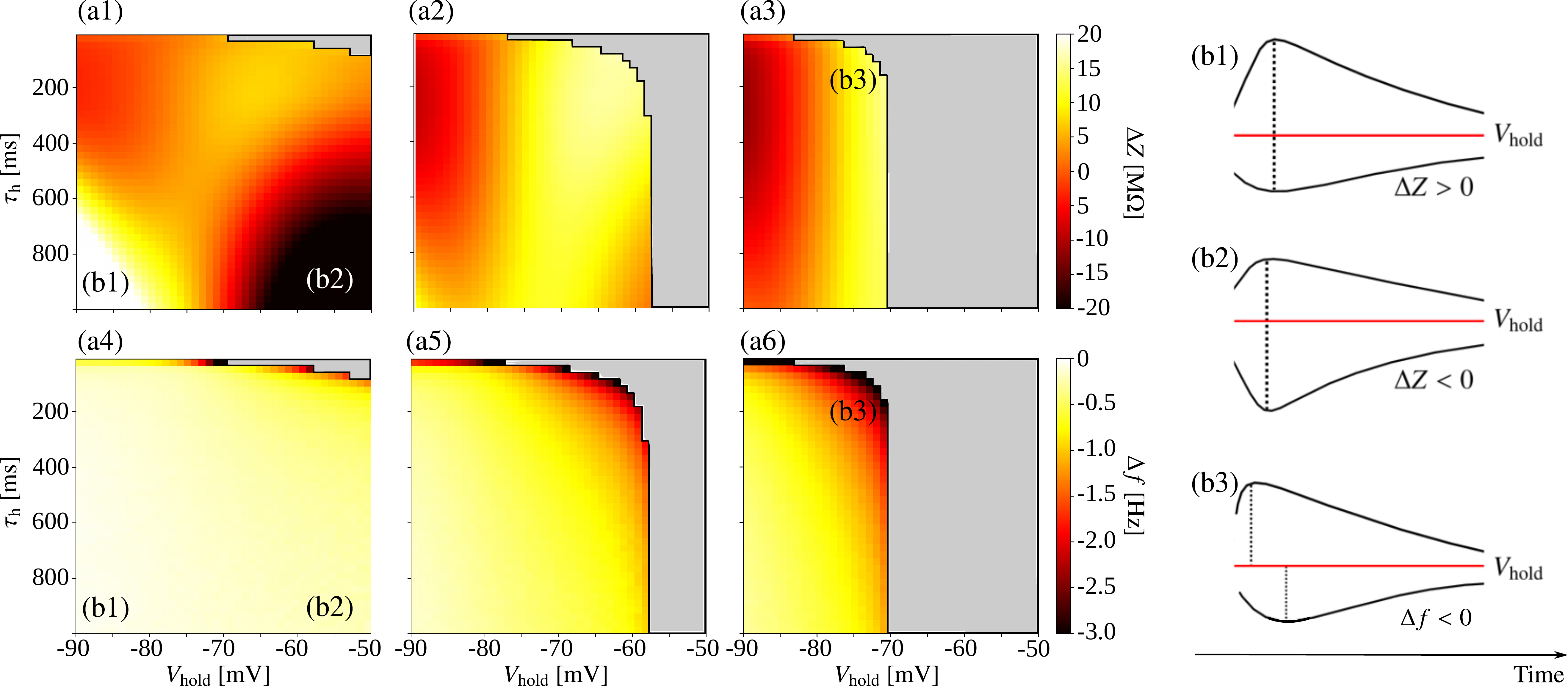}
	\caption{$V_{\rm hold}$--$\tau_{\rm h}$ diagrams for the $\Delta Z$ and $\Delta f$ shifts. (a1--a3) $\Delta Z$ for different values of $A_{\rm in}$. (a4--a6) $\Delta f$ for different values of $A_{\rm in}$. (a1,a4) $A_{\rm in}$=$0.1$ nA, (a2,a5) $A_{\rm in}$=$0.3$ nA, and (a3,a6) $A_{\rm in}$=$0.5$ nA. Light gray area represents excluded parameter regions due to the low-pass behavior of $Z^+$. (b1--b3) Schematic representations of the voltage envelopes for three selected points in the $V_{\rm hold}$--$\tau_{\rm h}$ diagram indicated by (b1,b2,b3). }
	\label{Fig::fig5}
\end{figure*}

\par Interestingly, the existence of resonance in $Z^{-}$ is not affected by the amplitude of the input current (Figs.~\ref{Fig::fig4}(a,b)). This might be due to the fact that $I_{\rm h}$ is activated by hyperpolarization. Since $I_{\rm h}$ is not activated by depolarization, the only remaining effects are due to the leak current making the response of the system linear. In addition, it is expected that for small $\tau_{\rm h}$ the system also behaves in a linear way \cite{rotstein2014}, which surely hampers the presence of non-linearities responsible for the asymmetries.

For a further characterization of the resonance properties under high ZAP current amplitude, we quantified the differences between the depolarizing and the hyperpolarizing membrane resonance peaks and frequencies (Fig.~\ref{Fig::fig1A}). We calculated the difference between the resonance peaks as $\Delta Z = Z^{+}(f^{+}_{\rm res}) - Z^{-}(f^{-}_{\rm res})$ and the shift between the resonance frequencies as $\Delta f= f^{+}_{\rm res} - f^{-}_{\rm res}$. We show these shifts for different combinations of $\tau_{\rm h}$ and $V_{\rm hold}$ in Fig.~\ref{Fig::fig5}. Gray area indicates that $Z^+$ has no resonance.

\begin{figure*}[!htb]
	\centering
\includegraphics[scale=0.4]{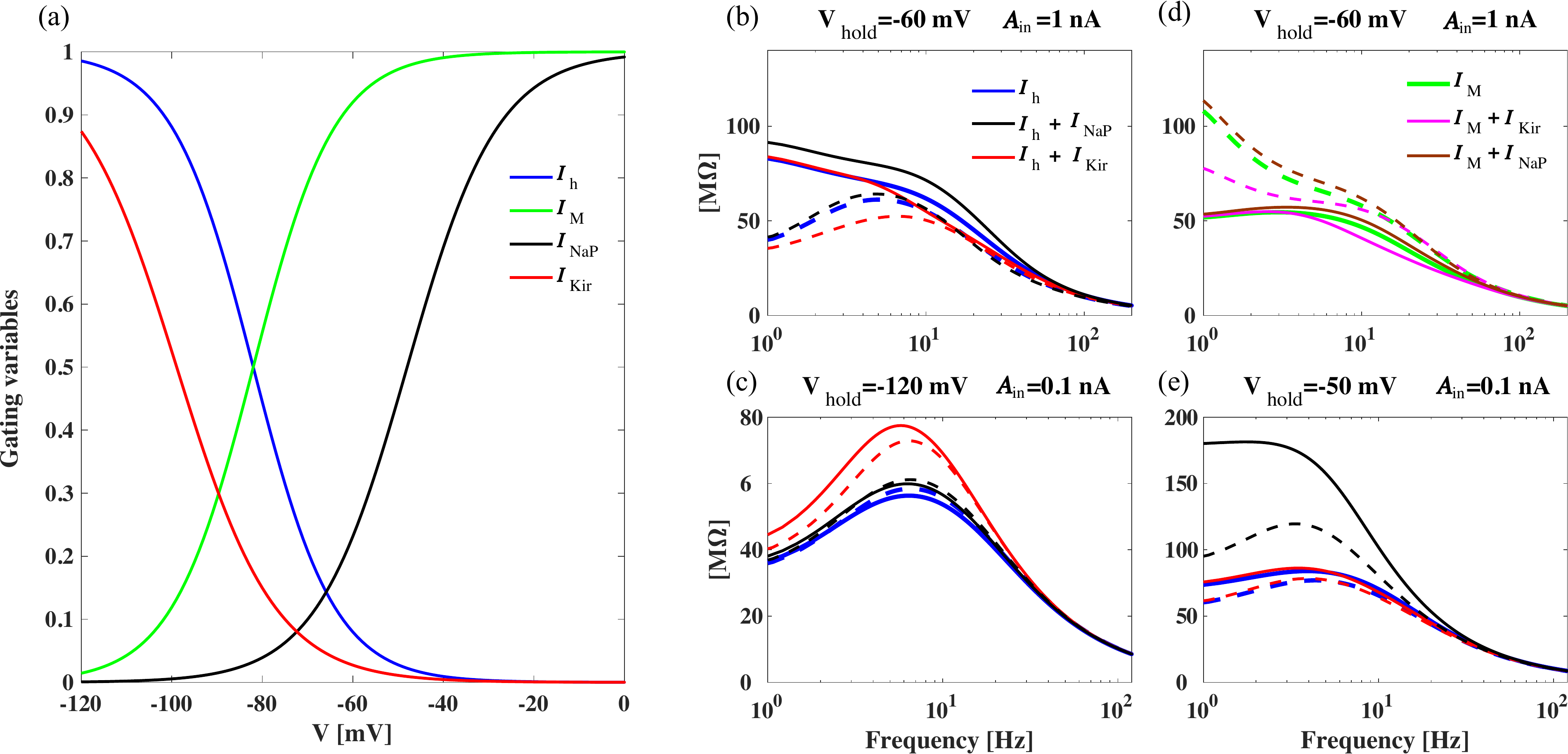}
	\caption{$Z^{+}$ and $Z^{-}$ behavior for other types of currents. (a) Activation curves for the currents that are tested. (b--e) $Z^{+}$ and $Z^{-}$ in solid and dashed lines, respectively. The currents in each simulation follow the same colors as in (b), the only difference is in (d) where the currents are presented in the legend. $V_{\rm hold}$ and $A_{\rm in}$ values atop.}
	\label{Fig::figCurrents}
\end{figure*}

\par Different combinations of $\tau_{\rm h}$ and $V_\text{\rm hold}$ can modulate the neuron's response in both amplitude and frequency (Fig.~\ref{Fig::fig5}). For small amplitudes such as $A_{\rm in}$=$0.1$~nA, the kinetics of $I_{\rm h}$ controls the effect of $V_{\rm hold}$ on $\Delta Z$ (Fig.~\ref{Fig::fig5}(a1)). When $\tau_{\rm h}$ is high we see two effects: whereas hyperpolarized $V_{\rm hold}$ amplifies $Z^+$, depolarized $V_{\rm hold}$ amplifies $Z^-$ (see the schemes for points (b1) and (b2) in Fig.~\ref{Fig::fig5}). On the other hand, lower $\tau_{\rm h}$ values weaken the $\Delta Z$ sensitivity to $V_{\rm hold}$. As $A_{\rm in}$ increases, the influence of $\tau_{\rm h}$ on $\Delta Z$ decreases, which may be related to the lack of activation of $I_{\rm h}$ at high amplitudes. In other words, this latter observation is related to higher amplitudes resulting in large displacements of the membrane potential consequently masking the $V_{\rm hold}$ effect. These remarks suggest a high sensitivity of the impedance amplitude to intermediate current amplitudes ($A_{\rm in}$ = $0.1$~nA) and a saturation for higher amplitude currents ($A_{\rm in}$ $\geq 0.3$ nA).

In Fig.~\ref{Fig::fig5}(a4), $V_{\rm hold}$ has no apparent influence on $\Delta f$ for slow $\tau_{\rm h}$ but higher frequency shifts are displayed at fast kinetics, which increase with $V_{\rm hold}$. When we observe the panels with higher ZAP current amplitudes (Fig.~\ref{Fig::fig5}(a5,a6)),  $V_{\rm hold}$ begins to influence $\Delta f$  even for low $\tau_{\rm h}$. In general, an increase in the input amplitude corresponds to an increase in the absolute value of $\Delta f$. For regions close to  $Z^{+}$ with low-pass filtering behavior (light gray area), $f_{\rm res}^{+}$ is near zero and, consequently, $\Delta f$ is maximized. These effects can be visualized in the schematic representation for point (b3) in Fig.~\ref{Fig::fig5}.

Neuronal resonance generated by $I_{\rm h}$ is  governed by the interplay between the $I_{\rm h}$ kinetics and the voltage-dependent conductance. Our results here suggest that asymmetries in the depolarized and hyperpolarized resonances ($\Delta Z>0$, $\Delta Z<0$ and $\Delta f<0$) are also driven by the same two parameters, but mostly at high amplitudes of oscillatory inputs. More specifically, when the neuron is subjected to an oscillatory stimulation with amplitudes of the order of tens of pA, larger displacements of the membrane voltage in relation to $V_{\rm hold}$ are generated, and, consequently, $I_{\rm h}$ will be less activated for $V>V_{\rm hold}$. When $I_{\rm h}$ is insufficiently activated, the effect of $I_{\rm leak}$ dominates and the neuron behaves as a low-pass filter, which does not display resonance. This effect explains why higher $V_{\rm hold}$ values abolish $Z^{+}$ resonance whereas $Z^{-}$ resonance remains. Furthermore, we predict that currents with opposed monotonic behavior on the activation curves with respect to the voltage (activation by hyperpolarization) could reverse this phenomenon, i.e. currents with activation for depolarized voltage values would generate resonance in $Z^{+}$ but not in $Z^{-}$ (see for example resonance exclusively on upper voltage envelopes at Figs.~6~A2 and A3 in Ref.~\onlinecite{schreiber2009} where $Z^{+}$ would demonstrate such effect).

\subsection{Effect of activation curves on asymmetries}\label{sec:currents}

In this subsection we explore the effect of other ionic currents with different combinations of activation curves and reversal potentials. We chose $I_{\rm M}$ which is again a resonant current but now activated by depolarization instead of by hyperpolarization as $I_{\rm h}$ (Fig.~\ref{Fig::figCurrents}(a)). We also chose two amplifying currents, namely $I_{\rm NaP}$ and $I_{\rm Kir}$, which are activated by depolarization and activated by hyperpolarization respectively  (Fig.~\ref{Fig::figCurrents}(a)). Our hypothesis is that the shape of the activation curves of a given ionic current controls the asymmetrical behavior in $Z^{+}$ and in $Z^{-}$. In Fig.~\ref{Fig::figCurrents}(a) we show the activation curves of the ionic currents studied in this paper. A second hypothesis would be that the main responsible for the asymmetry is not the activation curve but its effect on the phase-plane of the system. If this is the case, one would obtain similar results by having nonlinearities stressed on the voltage-nullcline instead. We will check this second hypothesis in Subsection \ref{sec:v_null}.

Figs.~\ref{Fig::figCurrents}(b,d) show $Z^+$ and $Z^-$ in the case of large ZAP current amplitudes $A_{\rm in}=1$~nA for the model with $I_{\rm h}$ and for the model with $I_{\rm M}$, respectively. The $Z^{+}$ ($Z^{-}$) curves for one current are nearly mirrored by the $Z^{-}$ ($Z^{+}$) curves for the other current. Whereas for  $I_{\rm h}$ (Fig.~\ref{Fig::figCurrents}(b)) resonance occurs in the bottom envelopes, for $I_{\rm M}$ (Fig.~\ref{Fig::figCurrents}(d)) resonance occurs only in the upper envelopes. The addition of an amplifying current such as $I_{\rm Kir}$ and $I_{\rm NaP}$ in the models has little effect and slightly shifts $Z^{+}$ and $Z^{-}$, as observed in the figures.

In Figs.~\ref{Fig::figCurrents}(c,e) we show the amplification of the impedance due to $I_{\rm Kir}$ and $I_{\rm NaP}$ currents. While $I_{\rm Kir}$ preferentially amplifies at voltages close to $V_{\rm hold}=-120$~mV, $I_{\rm NaP}$ amplifies at $V_{\rm hold}=-50$~mV. We also note that the amplification is different in these cases, the latter mostly amplifies the voltage at zero frequency while the former amplifies both the voltage at a non-zero frequency (resonance amplification). This voltage-dependent amplification is a consequence of the activation curves (as depicted in Fig~\ref{Fig::figCurrents}(a)) and it has been demonstrated elsewhere \cite{rotstein2017n}.

\subsection{Phase-plane analysis characterization of asymmetries}

\begin{figure*}[!htb]
	\centering
\includegraphics[height=10cm,keepaspectratio]{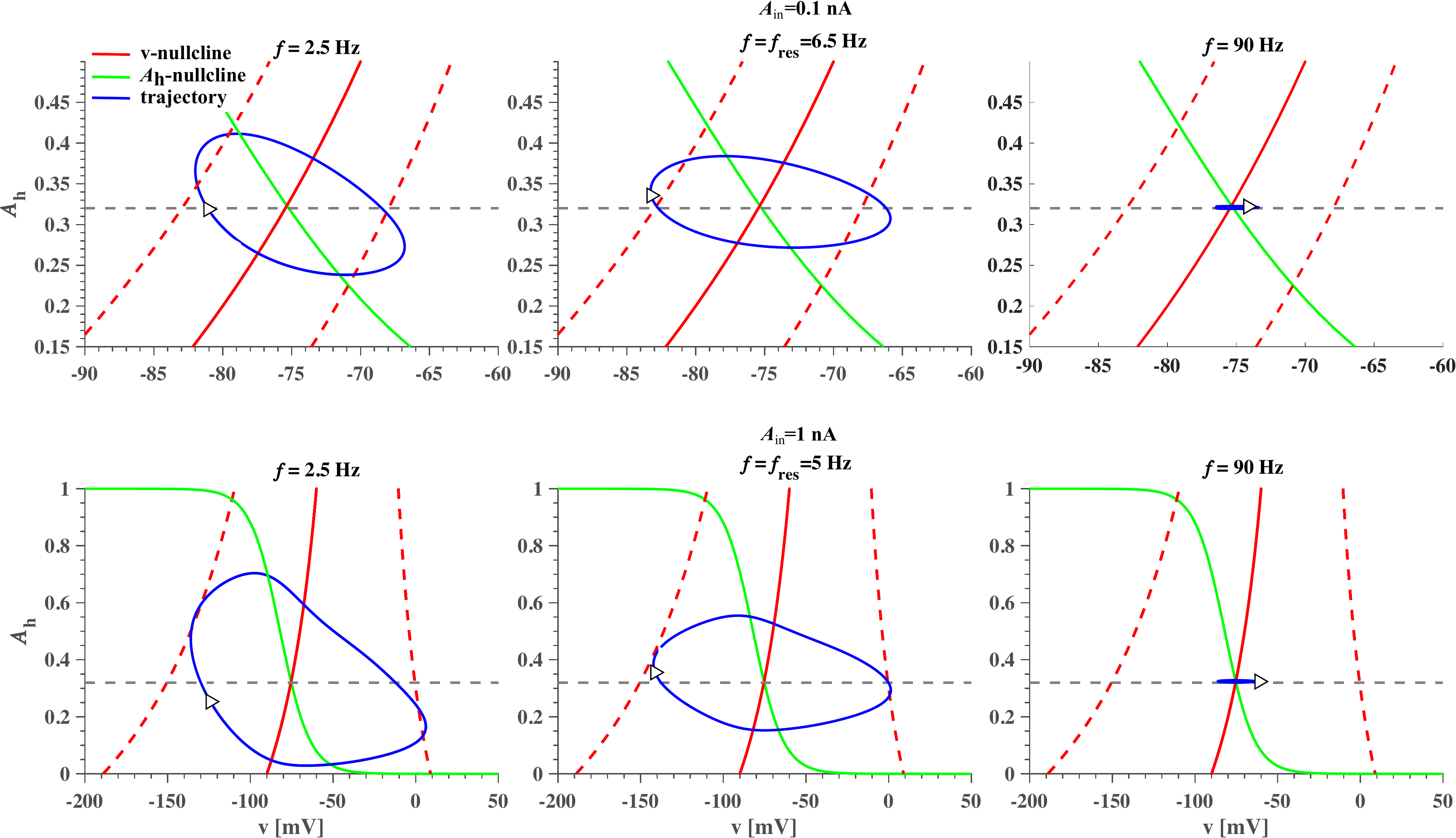}
	\caption{Phase-plane diagrams for the system with $I_{h}$ current in Fig.~\ref{Fig::figCurrents}(b). Trajectories for selected frequencies as indicated atop. Top row: $A_{\rm in}= 0.1$~nA. Bottom row: $A_{\rm in}=1$~nA. Blue:  Trajectory. Red line: $V$-nullcline for $I=0$~nA. Dashed red line: $V$-nullclines for $I=\pm A_{\rm in}$. Green line: $I_{\rm h}$-nullcline. Dashed gray line: intersection of nullclines at $I=0$.}
	\label{Fig::figIh}
\end{figure*}

\begin{figure*}[!htb]
	\centering
\includegraphics[height=9cm,keepaspectratio]{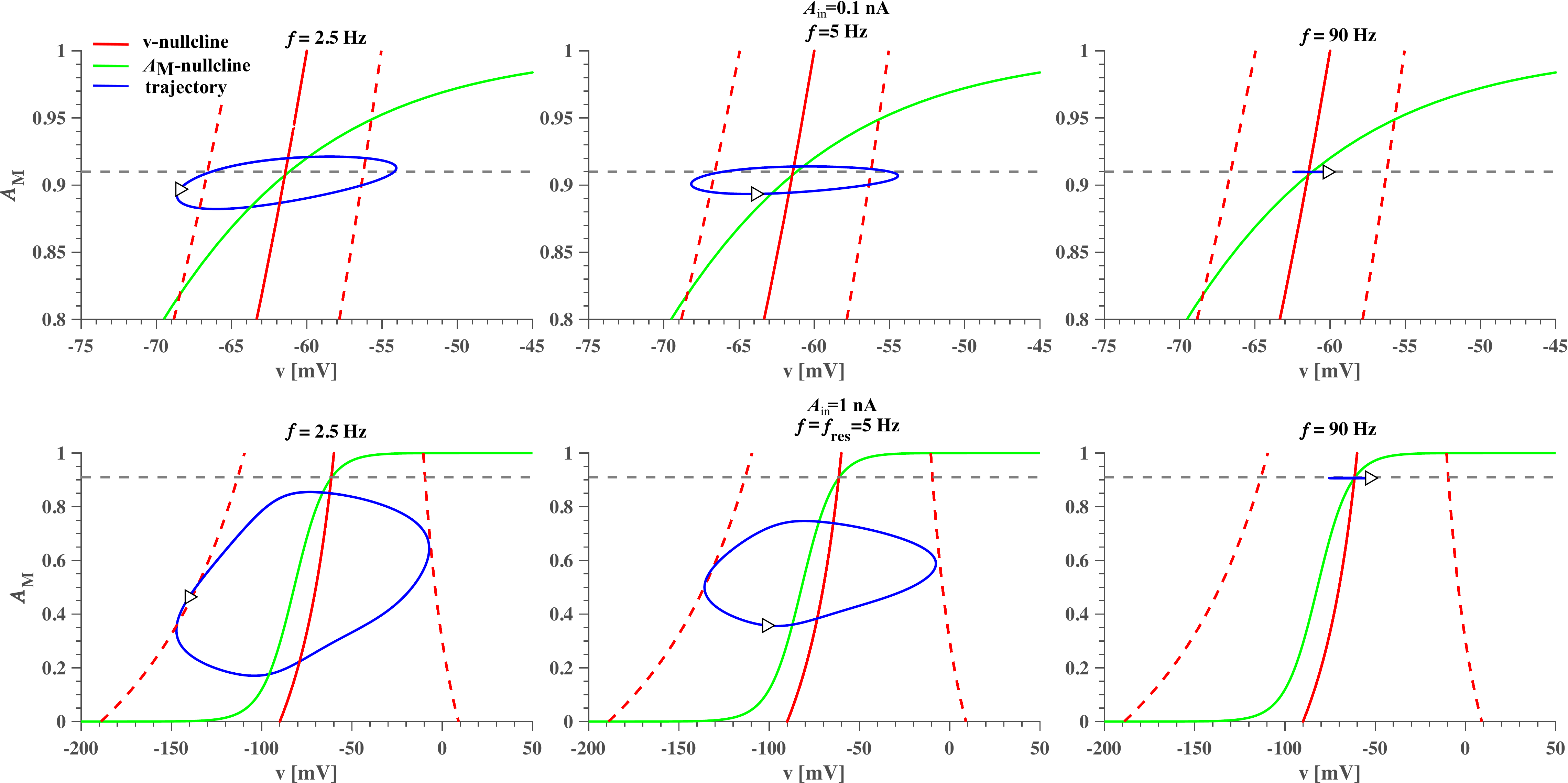}
	\caption{Phase-plane diagrams for the system with $I_{M}$ current in Fig.~\ref{Fig::figCurrents}(d). Top row: Trajectories for selected frequencies for amplitude $A_{\rm in} = 0.1$~nA. Bottom row: Same frequencies as above but with $A_{\rm in}=1$~nA. Blue:  Trajectory. Red line: $V$-nullcline for $I=0$~nA. Dashed red line: $V$-nullclines for $I=\pm A_{\rm in}$. Green line: $I_{\rm h}$-nullcline. Dashed gray line: intersection of nullclines at $I=0$.}
	\label{Fig::figIM}
\end{figure*}

In this subsection, we characterize the observations above using phase-plane analysis. By using such an approach we can provide the reader a geometrical understanding of how the voltage dynamics are affected by the ionic currents, thus allowing to explain and predict these dynamics. We present our results in Fig.~\ref{Fig::figIh} ($I_{\rm h}$)  and Fig.~\ref{Fig::figIM} ($I_{\rm M}$). We investigate the dynamics of the neuron in both the low and high amplitude cases in a frequency-dependent manner. We plot the nullclines either in red ($V$-nullcline for $dV/dt=0$) or in green ($A_i$-nullcline for $dA_i/dt=0$). For reference, we also plot (dashed-red) the location of the $V$-nullclines corresponding to the peak and trough values of the oscillatory input and in dashed gray the intersection of the nullclines. Note that in these cases, so that we could show frozen trajectories, we substitute the ZAP current by a sinusoidal input current of a single frequency, which we can control. In other words, this method is equivalent to using the ZAP current as in Eq.~(\ref{eq:ZAP}) but for $F_{\rm start}=F_{\rm stop}$, and therefore it delivers the same output response with the advantage of observing it without the time-dependent input frequency (see Fig.~13 in Ref.~\onlinecite{rotstein2015} and Fig.~10 in 
Ref.~\onlinecite{rotstein2017n} for examples and details of this method).

We follow Ref.~\onlinecite{rotstein2014b} and describe our observations as below: at low frequencies (first panel; $f=2.5$ Hz) the trajectories follow the $A_{\rm h}$-nullcline.  Although we do not explore the limit case when $f\xrightarrow{}0$, this is a situation where the system stays in a point. As frequency increases, there is an observable rotation. The resonant frequency is observed when the space covered by the trajectory in the  voltage domain is maximal (middle panel; $f=6.5$ Hz). For even higher frequencies, there is an observable shrinkage in the limit cycle trajectory up to the point that only a small line is observed at very high frequencies (third panel; $f=90$ Hz). For such high frequencies the system becomes quasi-one dimensional and, in the limit $f\xrightarrow{}\infty$, the trajectory coincides with the fixed-point and the system lies in a point \cite{rotstein2015}. 

In the second row, we see how the system behaves for high amplitudes. Given that the trajectory covers a larger voltage range the neuronal dynamics is more exposed to nonlinearities on the plane. Nonlinearities on the plane shape the neuron’s response, which differs from the ones described above (small input amplitudes). From low to high frequencies (panels from left to right; $f=2.5$ Hz,  $f=5$ Hz, and  $f=90$ Hz) there are noticeable differences in voltage peaks and troughs which are captured by the shape of the trajectory. The amplitude of the peaks only become smaller for increasing frequency. However, the amplitude of troughs increase up to a peak at the resonant frequency, and then decrease. This effect is related to the resonance being exclusively found on the bottom envelope in this setup (compare to blue curves in Fig.~\ref{Fig::figCurrents}(b)). 

The mechanism of resonance in $Z^-$ and absence of resonance in $Z^+$ for $A_{\rm in}=1$~nA can be geometrically explained as follows. Note that the trajectories for low frequencies do not develop further than the dashed-red lines and that the green solid line intersects the leftmost red-dashed line creating an enclosed region. Due to this region, for low frequencies the troughs cannot develop with high amplitudes. As frequency increases a rotation is observed, the trajectory escapes from this region and troughs increase in amplitude until a maximal value (resonance). For higher frequencies we observe shrinkage because the system tends to quasi-one dimensional.

The same mechanism described above is present in Fig.~\ref{Fig::figIM}, but now since the ionic-nullcline is mirrored we see a resonance exclusively on the upper voltage envelope (see $Z^{+/-}$ examples in Fig.~\ref{Fig::figCurrents}(d)). For low amplitudes (first row) the trajectories follow the $A_{\rm M}$-nullcline and, as the frequency increases, they rotate and shrink. For higher amplitudes (second row), the troughs decrease monotonically and the peaks resonate. The shape of the $A_{\rm M}$-nullcline and how it crosses with the $V$-nullcline defines this behavior.

Yet, observations of the second row in Fig.~\ref{Fig::figIM} shows a surprising unexpected dependency of the activation variables with frequency. The trajectory does not move around the fixed point for low frequencies, but only does so for higher frequencies. To the best of our knowledge, such unexpected frequency-dependent pattern was not reported in other subthreshold resonance phase-plane analyses (see, e.g., Refs.~\onlinecite{izhikevich2003,rotstein2014,rotstein2014b,tchumatchenko2014,zhao2018}). This is a point often overlooked in different studies due to the fact that they rely on small amplitude currents. 

\begin{figure}[!htb]
	\centering
\includegraphics[height=7cm,keepaspectratio]{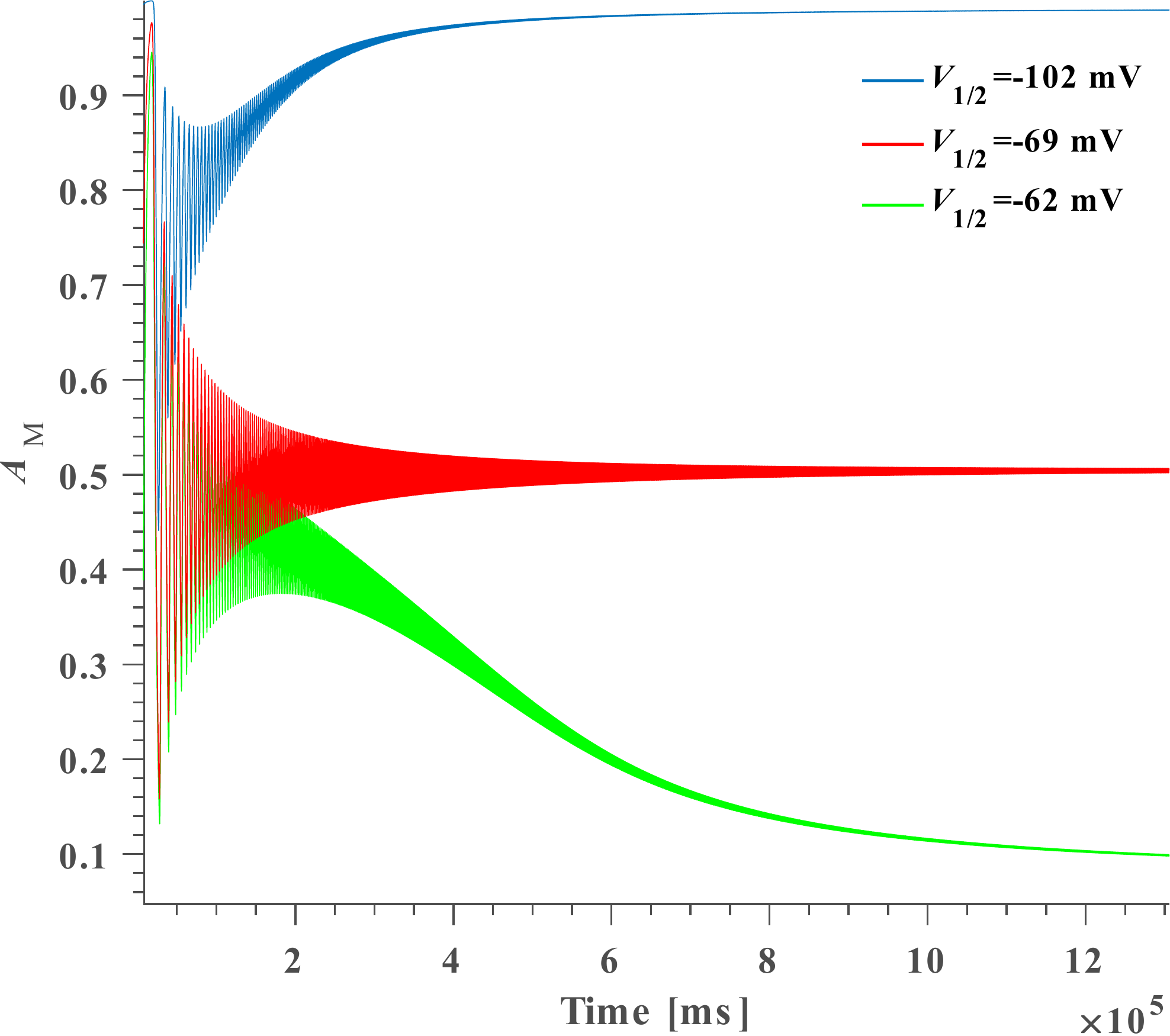}
	\caption{Frequency-dependent pattern of $A_{\rm M}$ variable upon ZAP stimulation for a neuron with only leak and $I_{\rm M}$ currents. The different colors represent three different values of the $V_{1/2}$ parameter as indicated in the legend.}
	\label{Fig::AM_curves}
\end{figure}

Our explanation of the above mentioned frequency-dependent pattern of $A_{\rm M}$ is threefold: it is first related to the proximity of the nullclines creating a region of slow velocity,  secondly to the time scale separation between $V$ and $I_{\rm M}$, and thirdly related to the activation properties of $I_{\rm M}$. As one can observe in the bottom row of Fig.~\ref{Fig::figIM}, the $V$-nullcline and the $A_{\rm M}$-nullcline are vertically close to each other creating a region with slow velocity in the proximity of these nullclines. Since there is a time scale separation, for low frequencies the trajectory follows preferentially the $V$-nullcline and, as frequency increases, it rotates and follows the $A_{\rm M}$-nullcline. However, in this situation, the $A_{\rm M}$-nullcline is not placed horizontally forcing the system to go through a transient rotation until it manages to reach the fixed point. Due to activation properties of $I_{\rm M}$ (see Fig.~\ref{Fig::figCurrents}(a)), the fixed point is reached through increase of $A_{\rm M}$. To investigate this behavior even further we explored in Fig.~\ref{Fig::AM_curves} the dynamics of the $A_{\rm M}$ activation variable while delivering the ZAP input to the neuron for a few parameters.

The frequency-dependent pattern of $A_{\rm M}$ can be controlled by adjusting the value of $V_{1/2}$. When we change $V_{1/2}$ we move the $A_{\rm M}$-nullcline and consequently the fixed point. If $V_{1/2}$ is close to the resting potential, then $A_{\rm M}$ is constant and it does not activate or deactivate with frequency; it is fixed at $A_{\rm M}=0.5$ such that $I_{\rm M}$ is half activated. For values of $V_{1/2}$ above (below) the resting potential, we found that the ionic channel will  activate (deactivate) towards the fixed point for increasing frequency. 

In the next section we will further explore the asymmetries in the voltage response using a simplified piecewise-linear model, and will provide a geometrical explanation of some of these phenomena.

\begin{figure*}[!htb]
	\centering
\includegraphics[height=8cm,keepaspectratio]{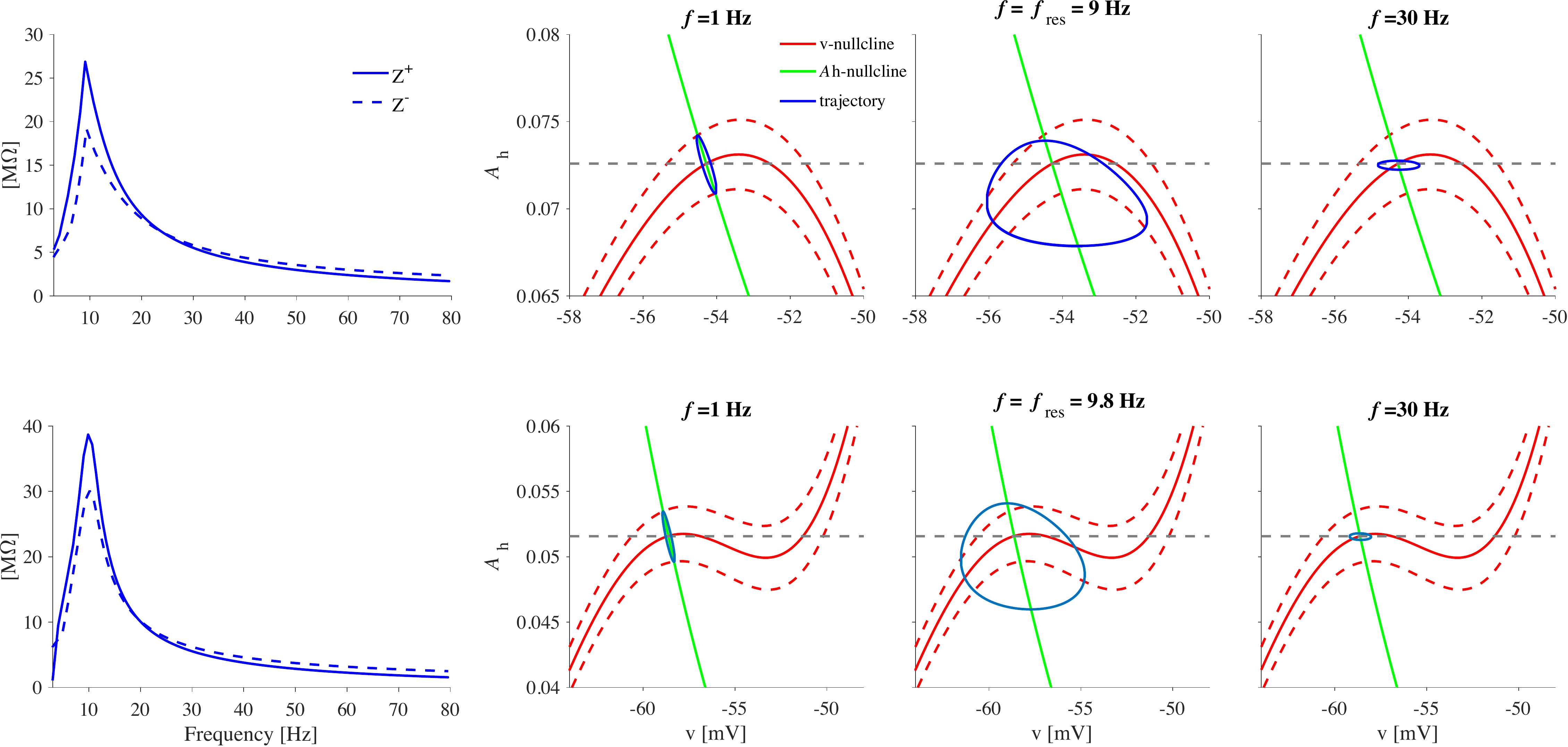}
	\caption{Asymmetrical response and phase-plane diagrams for the quadratic (top row) and cubic (bottom row) system. First column: $Z^+$ and $Z^-$ extracted from the systems upon ZAP current injection. Second to fourth column: trajectories for selected frequencies as indicated atop. Blue: Trajectory. Red line: $V$-nullcline for $I=0$~nA. Dashed red line: $V$-nullclines for $I=\pm A_{\rm in}$. Green line: $I_{\rm h}$-nullcline. Dashed gray line marks the intersection of the nullclines at $I=0$. In all cases $A_{\rm in}=0.1$.}
	\label{Fig::quadra_cubic}
\end{figure*}

\subsection{Asymmetries emerging from voltage nonlinearities}\label{sec:v_null}

In this subsection, we investigate if the asymmetrical response can still arise due to nonlinearities associated to the voltage variable and not associated to the ionic currents as we observed in the previous sections. We ask if the activation curves are the main reason for the observed asymmetries or if we could obtain qualitatively similar behavior by other means.  We chose conductance-based models from Ref.~\onlinecite{rotstein2017n} with $I_{\rm h} + I_{\rm NaP}$ currents. More specifically, we select model~1 and model~2 of that article. These models differ with respect to the parameters which introduce nonlinearities on the voltage-nullcline: it is either quadratic or cubic (see phase-planes in Fig.~\ref{Fig::quadra_cubic}).

\begin{figure}[!htb]
	\centering
\includegraphics[scale=0.3]{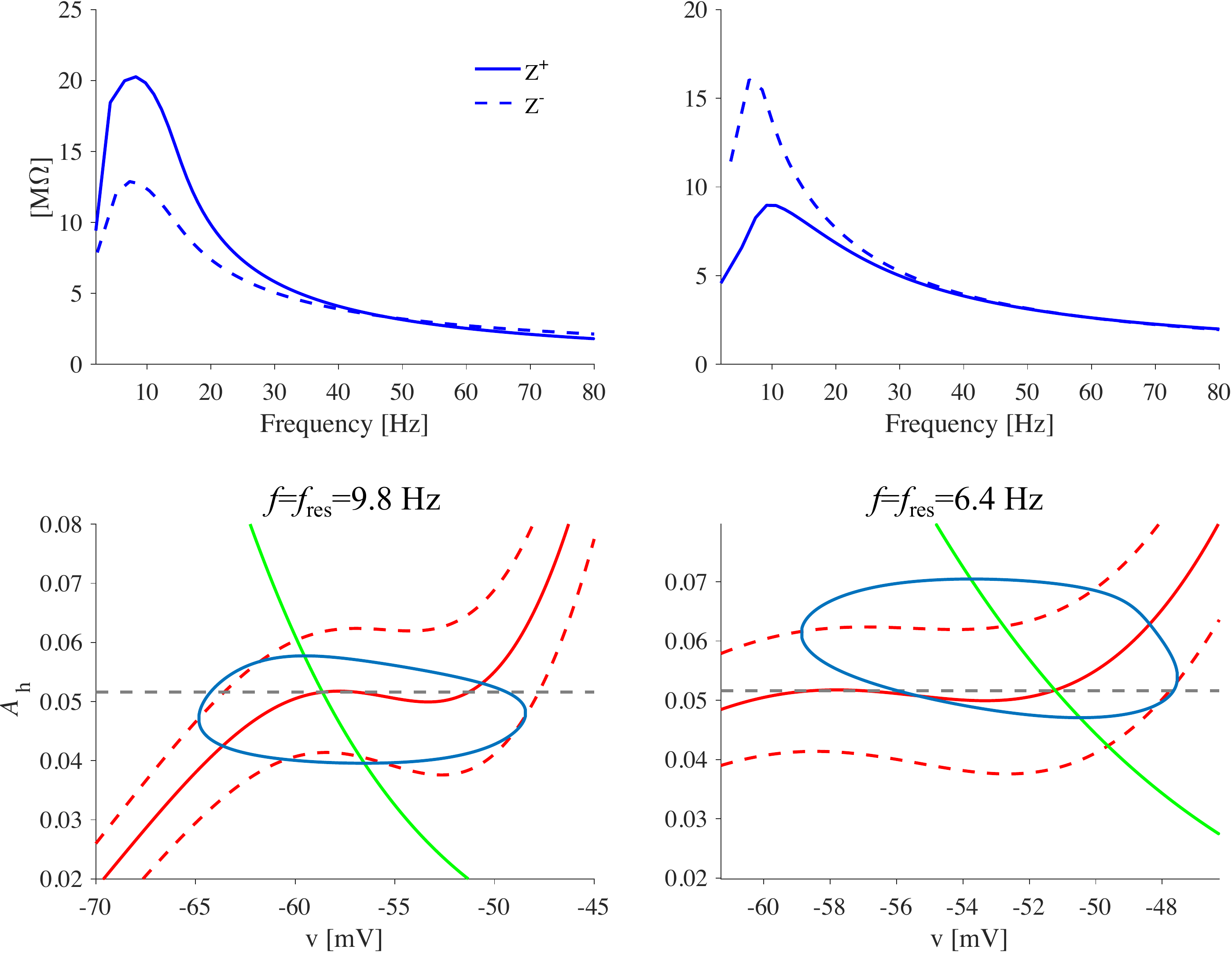}
	\caption{Asymmetrical response and phase-plane diagrams for the cubic model with higher ZAP current amplitude ($A_{\rm in}=0.5$). The case in the right had its $A_{\rm h}$-nullcline shifted to higher voltage levels for comparison.
	Top row: $Z^+$ and $Z^-$ extracted from the systems upon ZAP current injection. 
	Bottom row: trajectories for selected resonant frequency (see atop). 
	Red line: $V$-nullcline for $I=0$~nA. 
	Dashed red line: $V$-nullclines for $I=\pm A_{\rm in}$. 
	Green line: $I_{\rm h}$-nullcline. 
	Dashed gray line marks the intersection of the nullclines at $I=0$.}
	\label{Fig::cubic_high_amp}
\end{figure}

In comparison with the model that we have been using (Eqs.~(\ref{eq:HH}--\ref{eq::a_inf})), these models follow the same equations as in the other sections, but we emphasize two differences between them: their parameters and the fact that the $I_{\rm NaP}$ current has its activation variable approximated to its steady state as in Eq.~(\ref{Eq:model_vs}). Model 1 (quadratic) has the following parameters: For the leak current, we use $g_{\rm leak}=0.5$~mS/cm$^2$ and $E_{\rm leak}=-65$~mV; for $I_{\rm NaP}$, we use $\bar{g}_{\rm NaP}=0.5$~mS/cm$^2$,  $E_{\rm NaP}=55$~mV, $V_{1/2}=-38$~mV, $k=6.5$~mV, and $s=-1$; for $I_{\rm h}$, we use $\bar{g}_{\rm h}=1.5$~mS/cm$^2$, $E_{\rm h}=-20$~mV, $\tau_{\rm h}=80$~ms, $V_{1/2}=-79.2$~mV, $k=9.78$~mV, and $s=1$; for the external current, we use $I_{\rm DC}=-2.5$~$\mu A$/cm$^2$ and ZAP current with $A_{\rm in}=0.1$~$\mu A$/cm$^2$. Model 2 (cubic) has the following parameters: For the leak current, we use $g_{\rm leak}=0.3$~mS/cm$^2$ and $E_{\rm leak}=-75$~mV; for $I_{\rm NaP}$, we use $\bar{g}_{\rm NaP}=0.08$~mS/cm$^2$,  $E_{\rm NaP}=42$~mV, $V_{1/2}=-54.8$~mV, $k=4.4$~mV, and $s=-1$; for $I_{\rm h}$, we use $\bar{g}_{\rm h} = 1.5$~mS/cm$^2$, $E_{\rm h}=-26$~mV, $\tau_{\rm h}=80$~ms, $V_{1/2}=-74.2$~mV, $k=7.2$~mV, and $s=1$; for the external current, we use $I_{\rm DC}=0.01$~$\mu A$/cm$^2$ and ZAP current with $A_{\rm in}=0.1$~$\mu A$/cm$^2$. For a more detailed description of the model we refer the reader to 
Ref.~\onlinecite{rotstein2017n}.

\begin{align} 
  \label{Eq:model_vs}
  I_{N a p} &=\bar{g}_{\rm NaP} A_{\rm NaP}^{\infty}(V)\left(V-E_{\rm NaP}\right)
\end{align}

In Fig.~\ref{Fig::quadra_cubic}, we present a comparison of the quadratic and cubic models. In both models, the asymmetric response emerges. In the first column, we see that $Z^+ > Z^{-}$ and that the quadratic and cubic models show qualitatively the same behavior. Given that the $A_{\rm h}$-nullcline adds a quasi-linear influence on the trajectories, the quadratic/cubic $V$-nullcline is the only nonlinear influence on the response of the model. For $f\xrightarrow{}0$ and $f\xrightarrow{}\infty$, the trajectory follows the same behavior described above: at slow frequencies, it follows closely the $A_{\rm h}$-nullcline and at fast frequencies, it converges to the fixed-point. Resonance occurs at intermediate frequencies when the trajectory covers a larger voltage range.

As observed in the models from previous sections, the amplitude has a strong effect in the asymmetrical response. In the case of the quadratic model, however, if the ZAP current amplitude is enhanced action potentials develop. This shows a limitation to the increase of $\Delta Z$. Nonetheless, in the case of the cubic model a further increase of the ZAP current amplitude is possible because the trajectory will develop around the cubic $V$-nullcline. Note this effect in Fig.~\ref{Fig::cubic_high_amp} in the first column.

A second effect from the cubic $V$-nullcline concerns the position of the fixed-point which can be controlled by the $A_{\rm h}$-nullcline. We perform this study which is presented in the second column of Fig.~\ref{Fig::cubic_high_amp}. Clearly, there is an inverted behavior: now $Z^- > Z^+$.  

These experiments confirm that asymmetrical response may emerge either from ionic currents or from the voltage nullcline. But, most importantly, it is a phenomenon which emerges from nonlinear systems.

\subsection{Characterization in a reduced system}\label{sec:geom}

So far, we have shown two different, but related phenomena: (i) asymmetric responses that are a property of nonlinear systems, and that may or may not be pronounced. The asymmetries become pronounced mostly as a function of the amplitude of the oscillatory input, the time scale separation or holding voltage; (ii) the frequency-dependent pattern of the gating variable, which is related to a combination of the voltage and ionic current properties.

In this subsection, we will use a simplified model to geometrically test and explain (i), but we are not able to explore (ii) because this model does not contain any activation dynamics. First, we identify what are the most important parameters that control the asymmetric responses and how the latter depends on the input amplitude. For simplicity, we chose a piecewise-linear model (PWL) to explain the asymmetry through a geometrical approach on the phase-plane \cite{rotstein2014b}. Nonlinearities are added by a sudden change in the slope of one of the nullclines. In this respect, we believe that the PWL system is simple enough to be addressed as a reduced system. The PWL system reads

\begin{align}
\frac{dv}{dt} & = h_v(v) - w + I_{\rm ZAP} \label{Eq:dv_pwl} \\
\frac{dw}{dt} & = \epsilon \left[h_w(v)-w + \beta \right],
\label{Eq:dw_pwl}
\end{align}

\noindent where the functions $h_v(v)$ and $h_w(v)$ are described by 

\begin{align}
h_v(v) & = 
\begin{cases}
\eta v & \text{ if } v \leq 0.2\\
\eta_r v & \text{ if } v > 0.2
\end{cases}
& & \text{  and } & h_w(v) = \alpha v,
\end{align}

\noindent where $\alpha=0.4$, $\eta=-1$, and $\eta_r=-0.2$. The values of $\epsilon$ (time scale separation) and $\beta$ (capturing the bias DC current) we use in our simulations will be indicated as needed.

\begin{figure}[!htb]
	\centering
\includegraphics[height=12cm,keepaspectratio]{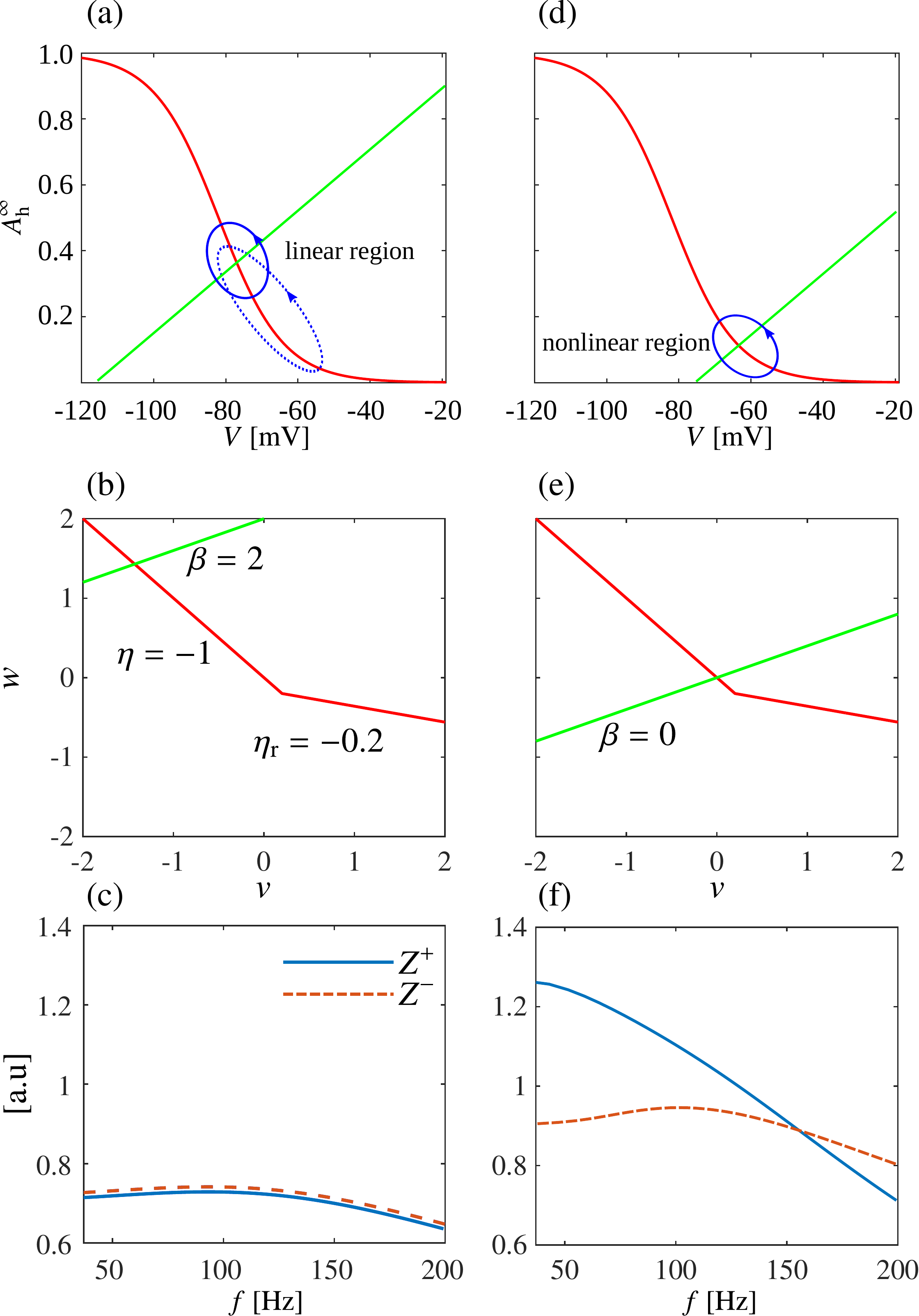}
	\caption{Approximation of the biophysical model with the PWL system. (a) Activation curve showing the different regions (approximately linear and nonlinear) achieved by moving the nullcline through $\beta$. (b) Nullclines of the PWL system with parameters shown in the panel. $\beta=2$, $\epsilon=1.0$, and ZAP current amplitude $A_{\rm in}=1.2$. (e) Nullclines for $\beta=0$. (c) $Z^+$ and $Z^-$ for the system in (b). (f) $Z^+$ and $Z^-$ for the system in (e).}
	\label{Fig::fig6}
\end{figure}

As we have shown in the previous section, $V_{\rm hold}$ plays a relevant role in determining the asymmetric response. Thus, here we search for parameters in the PWL system that represent changes in $V_{\rm hold}$. As one can observe from the schematic $I_{\rm h}$ activation curve in Fig.~\ref{Fig::fig6}(a), which mimics the $A_{\rm h}$ curve, as $V_{\rm hold}$ increases the $V$-nullcline gets closer to a region on the plane where nonlinearities are more prominent. We make a distinction between ``linear region'' and ``nonlinear region''. The areas where asymmetries are not present are denoted as ``linear regions'', and are found in the middle of the activation curve. The areas where asymmetries are present are denoted as ``nonlinear regions'', and are found where nonlinearities are more prominent on the plane. These regions can be mimicked in the PWL system by the breaking point that separates the two linear pieces. In this regard, in the phase planes of the PWL system in Figs.~\ref{Fig::fig6}(b,e) we have observed that as the nullclines are shifted by changing the values of $\beta$, the breaking point ($v=0.2$) becomes more distant or closer, the latter one emphasizes asymmetries because trajectories get closer to the ``nonlinear region'' (Figs.~\ref{Fig::fig6}(e,f)). In fact, we show that this is a general phenomenon, not only observed in a biophysical neuron model but also found in systems where nullclines are placed in such a way that they allow a certain competition between linear and nonlinear effects, i.e. between a ``linear region'' and a ``nonlinear region''. 

\begin{figure}[!htb]
	\centering
\includegraphics[height=12cm,keepaspectratio]{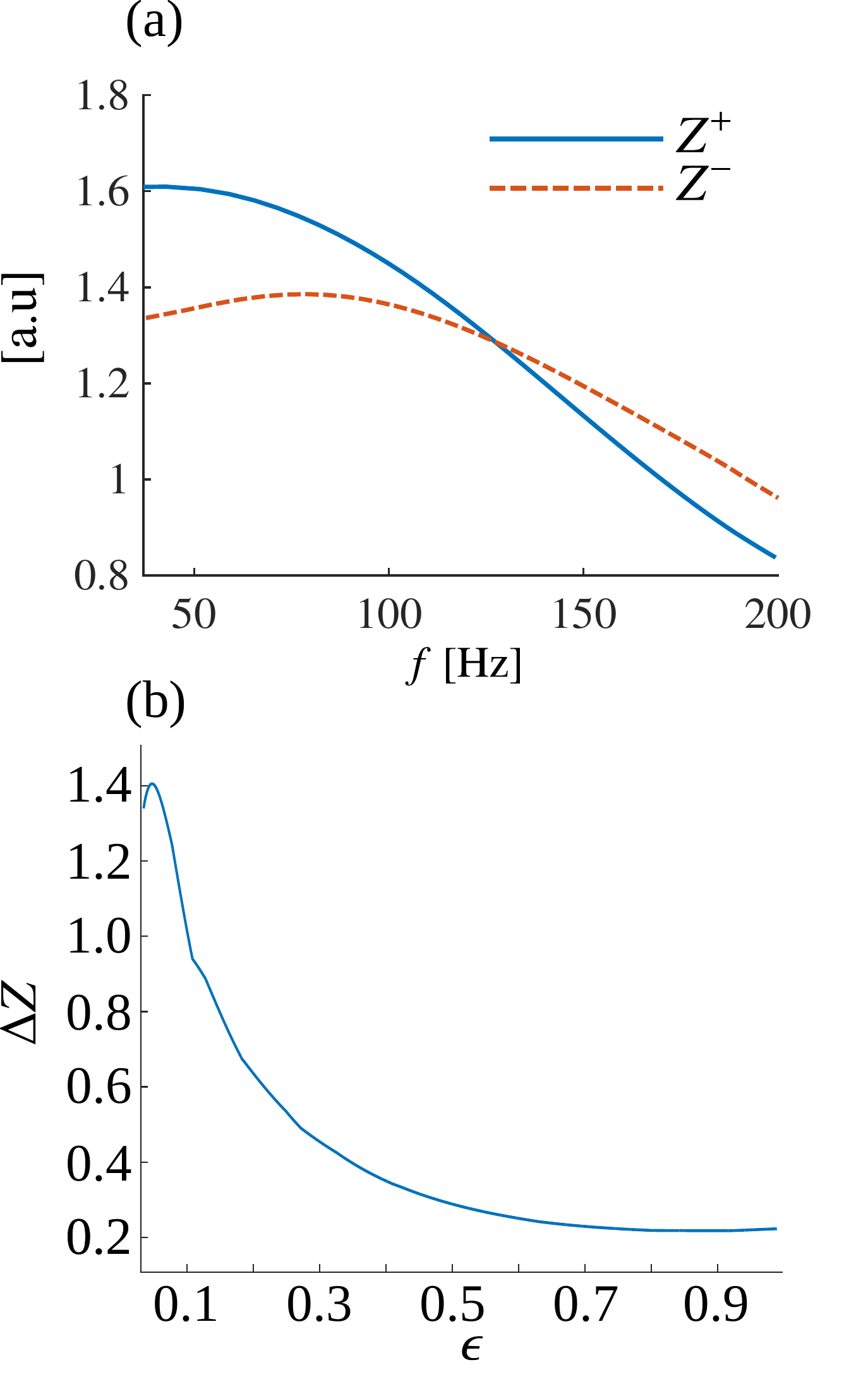}
	\caption{Dependence of amplitude and $\epsilon$ on the PWL. Parameters follow the same setup as in Fig.~\ref{Fig::fig6}(b) and (e) but with amplitude $A_{\rm in}=3$. (a) $Z^+$ and $Z^-$ for higher amplitude. (b) Dependency of $\Delta Z$ with $\epsilon$.}
	\label{Fig::fig7}
\end{figure}

Nonetheless, if asymmetries are created by the fact that oscillatory trajectories lie around the non-linearities in the $I_{\rm h}$ activation curve (or equivalently, close to the breaking point in the PWL system), other mechanisms could achieve the same effect. Here, we were able to retrieve asymmetries by increasing the ZAP current amplitude which forces the trajectories of PWL system to the nonlinear region again (see dotted trajectory in the schematic representation in Fig.~\ref{Fig::fig6}(a)). This simulation uses the same system of Figs.~\ref{Fig::fig6}(b,c) for higher amplitude and is presented in Fig.~\ref{Fig::fig7}(a) where we retrieved the asymmetries.

Further we analyzed the effect of varying the time-scale separation $\epsilon$ in the PWL system (see Section~\ref{sect:dependencies} where we discuss time-scale separation controlled by $\tau$).
In addition, as previously reported \cite{rotstein2014b}, an increase in the time-scale separation plays an important role in amplifying the voltage response. We tested the time-scale separation with $\epsilon$ because it corresponds to the inverse of the time-scale of the system. Our simulations confirm that resonance emerges for $0 < \epsilon < 1$ and is more pronounced the smaller the value of $\epsilon$ within some range (similar results can be found in Ref.~\onlinecite{rotstein2014}). At the same point, the asymmetries observed show maximum of $\Delta Z$ (see Fig.~\ref{Fig::fig7}(b)). 

\begin{figure}[!htb]
	\centering
\includegraphics[scale=0.35]{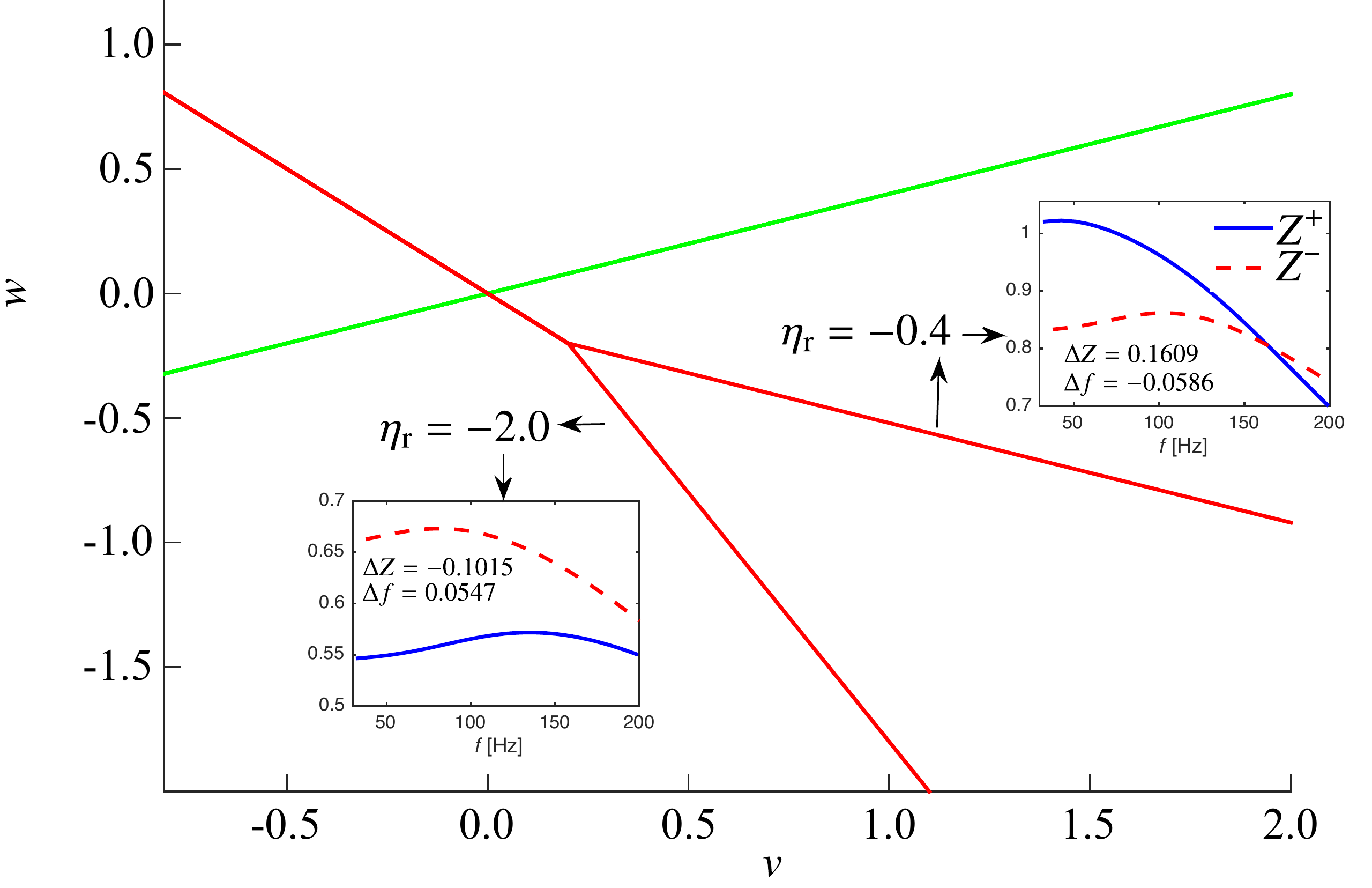}
	\caption{PWL nullclines for two different values of $\eta_{\rm r}$ as indicated by the arrows. Insets: $\Delta Z$ and $\Delta f$ for the two values of $\eta_{\rm r}$.}
	\label{Fig::fig8}
\end{figure}

The results using the PWL system show an asymmetrical response emerging close to the breaking point, but we remind the reader that the $I_{\rm h}$ activation curve posses two nonlinear regions: one for hyperpolarized $V_{\rm hold}$ and another one for depolarized $V_{\rm hold}$. At the same time Fig.~\ref{Fig::fig5} shows that $\Delta Z$ exhibits an inversion of signals at $V_{\rm hold} \approx 82$ mV, which could be attributed to the trajectory of the system being closer to another nonlinear region of the curve. To understand this behavior we used the PWL system where the asymmetrical response was evaluated under $\eta_r$ variation in order to simulate both extremes of the $I_{\rm h}$ activation curve. Note that for $\eta_r<-1.0$ ($\eta_r>-1.0$) the PWL system is comparable with the already mentioned hyperpolarized (depolarized) nonlinear portion of the $I_{\rm h}$ activation curve. This is demonstrated in Fig.~\ref{Fig::fig8} where we show how different angles determined by $\eta_r$ can influence $\Delta Z$. 

The PWL system could be used to describe other asymmetrical properties. An example is found in Ref.~\onlinecite{schreiber2009} where resonance was found only in the upper envelope and not in the bottom envelope. As we have shown in the previous section, this is related to activation curves such as $I_{\rm M}$ which activates with respect to the voltage instead of the deactivation observed for example in $I_{\rm h}$ (see Fig.~\ref{Fig::figCurrents}(a)).  By working with the nullclines extracted from Eqs.~(\ref{Eq:dv_pwl}--\ref{Eq:dw_pwl}) and producing phase-planes with qualitatively mirrored images we were able to recover situations where resonance is only found in the upper envelope confirming that the simple PWL system is a good geometrical approximation to generate asymmetrical responses \cite{rotstein2014b}.

\section{Discussion}

Upon receiving an oscillatory input with a varying frequency, a neuron is said to exhibit resonance when the voltage response peaks at a non-zero input frequency. In general, the oscillatory input is injected with relatively low amplitude therefore producing a quasi-linear voltage response, which is almost symmetric with respect to the holding potential. In addition, a quasi-linear response  has its voltage number of cycles equals the number of input cycles. However, when the amplitude is increased, asymmetries in the voltage response envelope arise. Since the impedance profile is essentially an average of the upper and lower voltage response envelope profiles, in the presence of asymmetries, this impedance fails to capture important properties of these envelopes. Of particular importance is the case where the maximum of the upper envelope and the minimum of the lower envelope  occur at different frequencies. In these cases, the impedance and the upper envelope predict different characteristic frequencies at which the subthreshold responses are communicated optimally to the spiking regime. Although asymmetries in the voltage response to oscillatory inputs have been observed both theoretically and experimentally  \cite{beraneck2007,tohidi2009,fischer2018}, their properties and the mechanisms that give rise to these asymmetric envelope profiles have not received much attention. 
To our knowledge, this is the first systematic study that addresses these issues.

Understanding how the neuronal intrinsic properties control a given response  pattern can have implications for neuronal information processing in addition to an accurate prediction of the response spiking patterns to similar inputs (communication of oscillatory inputs to the spiking regime). One of them is for the post-inhibitory rebound (PIR) phenomenon by which inhibitory inputs give rise to spiking activity in an otherwise silent neuron. PIR and resonance have been associated in the literature \cite{hasselmo2014} since the currents that produce resonance (e.g., $I_{\rm h}$) are involved in PIR.  Experimental observations agree that network resonance can emerge by inhibitory communication by means of PIR \cite{stark2013}. While the precise link between these two phenomena is not known, we speculate that the characteristic time scale associated to the trough of the lower response envelope plays a role in determining the PIR time scale, and therefore sheds light on the cellular mechanisms controlling the communication of inhibitory inputs to the spiking regime. More research is necessary to develop these ideas in a more precise fashion.

We unfolded the impedance profile into two quantities, the upper and lower impedance profiles, by taking the absolute values of the differences between the upper/lower voltage envelopes and the holding potential and normalizing them by the input amplitude. 
These metrics allowed a further characterization of asymmetry properties in terms of the model parameters. 
More specifically, we manipulated the nonlinearities from three different perspectives: (i) biophysical, by looking at the different representative ionic currents ($I_{\rm h}$, $I_{\rm M}$, $I_{\rm NaP}$, and $I_{\rm Kir}$); (ii) nonlinear voltage dependencies, by looking at the type of nonlinearities present in the model as captured by the voltage-nullclines (quadratic, cubic); and (iii) nonlinear mechanisms, by looking at the role that the nonlinearities play in different voltage ranges by using a piecewise-linear model that allows for changes of the nonlinearity type in one region but not in  others, therefore allowing a more detailed analysis. 
This combined approach produced rules determining the effects that different types of ionic currents have on the upper and lower impedance profiles. These predictions  are amenable for experimental testing using the dynamic clamp technique \cite{sharp1993,prinz2004}. 

From the biophysical point of view we focus on how  different resonant ($I_{\rm h}$ and $I_{\rm M}$) and amplifying ($I_{\rm NaP}$ and $I_{\rm Kir}$) currents shape the upper/lower impedance profiles. We chose these four currents because they are representative of the four different type of scenarios. $I_{\rm NaP}$ and $I_{\rm M}$ are depolarization-activated, but $I_{\rm NaP}$ is inward, while $I_{\rm M}$ is outward \cite{izhikevich2007}. Similarly, $I_{\rm h}$ and $I_{\rm Kir}$ are hyperpolarization-activated, but $I_{\rm h}$ is inward, while $I_{\rm Kir}$ is outward \cite{izhikevich2007}. In particular, we characterized 
the effect that the holding potential has in determining the upper/lower impedance properties since the holding potential controls how much the neuron dynamics is affected by the activation curve of the currents. Depending on where the neuron's voltage is located, the ionic current has a higher effect either depolarizing or hyperpolarizing the neuron and eventually causing resonance or amplification on the upper or lower response envelope, respectively. Notably, the gating variables showed an unexpected frequency-dependent pattern which is related with the half-activation value of the sigmoid function: if the half-activation value is above or below the resting potential, the ionic current will be either activated or deactivated with increasing frequency, respectively. This effect is not completely understood and is probably related to strong nonlinearities associated to the gating variables. However, to the best of our knowledge, this is the first time such a phenomenon is discussed and it could  potentially elucidate new functions for the ionic currents. This calls for additional research.

From the dynamics point of view, we provided a geometrical interpretation of the role of the asymmetries in shaping the voltage responses using a dynamical systems analysis of both the biophysical model and a simplified piecewise-linear model. We show how the asymmetric responses are linked to the model nonlinearities, which are captured by the nullclines in the phase-plane diagram, and the time scale separation between the participating variables. We found that higher current amplitudes are more likely to generate asymmetric responses because the systems' trajectories are more prone to reflect nonlinearities from the plane since they cover larger areas of the phase-plane where the nullcline nonlinearities on both sides of the holding potential differ more than in a close vicinity of it. Moreover, by taking advantage of the ability to modify one branch of the voltage nullcline, while keeping the other intact in piecewise linear models, we learned that asymmetries exist even in a system without currents with kinetics (activation curves). Markedly, this result demonstrates that the gating variables are not essential to obtain asymmetrical responses as they naturally emerge in simplified systems as long as the input amplitude, the nullclines position, and the time-scale separation allow such effect. 

We have carried out simulations with longer ZAP and the same frequency range for representative parameter values and we have observed no difference with the results discussed in the paper (not shown). We have also carried out simulations in reverse frequency order (starting with the higher frequencies and ending with the lowest ones) and, again, we have observed no differences (not shown). This indicates that, as expected, there are no intrinsic cellular processes that are too slow as to fail to be captured by the ZAP currents we have used in this work.

Our results are in line with recent experimental observations \cite{tohidi2009,fischer2018}. In Ref.~\onlinecite{tohidi2009} for example, the authors extensively studied how different currents shape the upper and lower envelopes of the voltage profile in neurons in the crab pyloric CPG. Recently, a clear asymmetric subthreshold voltage response was experimentally recorded in Ref.~\onlinecite{beraneck2007} (see Figs.~3 and 4 of that article). There, by applying ZAP currents to frog vestibular neurons, the authors observed asymmetrical voltage responses where the resonant frequency of the upper envelope increases with depolarization, i.e. $\Delta f$ changes with $V_{\rm hold}$ as observed in our simulations. That work allowed a better classification of neurons in the frog vestibular system by means of their response properties.  Another experimental example where asymmetries were recorded is found in Ref.~\onlinecite{heys2010} where the disruption of M-currents with the selective blocker XE991 in medial entorhinal cortex cells mediated such pattern (see Fig.~7 of that article). The latter example indicates a clear channel dependency on the asymmetrical response of a neuron as we have reported here by the use of computational models. Another key point is that in all the models we have used, the number of output cycles coincide with the input cycles meaning that this linearity principle is maintained. Further research should be conducted to clarify impedance profiles in which this linearity principle is broken.

Our observations suggest important implications for signal processing in neurons. First, since the resonance peak difference $\Delta Z$ can be either negative or positive, we speculate that an oscillatory input to a neuron could lead it to preferably respond with a hyperpolarizing or a depolarizing displacement in the membrane potential. Second, the existence of the resonance frequency shift  $\Delta f$ suggests that the resonance properties of a neuron may be related to additional functions than increasing neuronal excitability. For $\Delta f \neq 0$, the system has two characteristic frequencies, one depolarizes, the other hyperpolarizes. However, we found the $\Delta f$ values to be not very large, which raises the question of whether there are other channel combinations, possibly including calcium concentration dependent channels, that could further increase $\Delta f$ or amplify its effect when the neuron is embedded in a network. 
 In this regard, we propose a role for neuronal response modulation under  oscillatory inputs for  different values of the $I_{\rm h}$ kinetics parameters and its voltage regulated activation. $I_{\rm h}$ displays a high variability with the time constant of its kinetics spanning from tens of milliseconds to several seconds \cite{poolos2006,jung2010,zemankovics2010,dougherty2013,ceballos2016}. This could be a potential mechanism for output modulation.

The characterization of $Z^{+}(f)$ and $Z^{-}(f)$ has implications for model parameter estimation of neuronal systems. When measuring the properties of ionic currents, the information about each one is usually obtained separately by means of experimental procedures (e.g., patch clamp, voltage clamp). However, these experiments do not provide information about the interaction between currents and it has been shown that the effects of these interactions cannot be understood in terms of the sum of the effects of the participating currents. In addition, because the number of model parameters are typically larger than the information obtained by applying constant pulses of current/voltage, the problem is unidentifiable (parameter values cannot be uniquely determined). Resonance experiments provide additional, independent dynamic information that can be used to help characterizing the ionic currents and their interactions, more so when the responses exhibit asymmetries.

As it has been shown, the impact of subthreshold resonance ranges from suprathreshold neuronal properties (e.g. firing rate) to network behavior and learning\cite{richardson2003,tchumatchenko2014,chen2016,roach2018}. Aligned with this, our results suggest mechanisms by which a network could switch between different states not only according to its activity and input frequency but also to hyperpolarization and depolarization. This can have implications on neuronal spiking properties in a wide range of circumstances.  These phenomena deserve further exploration.




\begin{acknowledgments}
 This paper was developed within the scope of the IRTG 1740 / TRP 2015/50122-0, funded by DFG / FAPESP. This work was partially supported by the Research, Innovation and Dissemination Center for Neuromathematics (FAPESP grant 2013/07699-0). This work was partially supported by the National Science Foundation grant DMS-1608077 (HGR, RFOP). RFOP was also supported by a FAPESP PhD scholarship (grant 2013/25667-8), CCC is supported by a CAPES PhD scholarship, VL is supported by a FAPESP MSc scholarship (grant 2017/05874-0), ROS is supported by a FAPESP PhD scholarship (grant 2017/07688-9) and ACR is partially supported by a CNPq fellowship (grant 306251/2014-0). This study was financed in part by the Coordena\c{c}\~ao de Aperfei\c{c}oamento de Pessoal de N\'ivel Superior - Brasil (CAPES) - Finance Code 001.
\end{acknowledgments}

\nocite{*}
%

\end{document}